\documentclass[useAMS,a4,usenatbib]{mn2e}
\input psfig.sty

\usepackage{floatflt, epsfig}
\usepackage{epstopdf}

\def \chisq  {\ifmmode  \chi^2   \else  $\chi^2$  \fi}  
\def \spose#1{\hbox  to 0pt{#1\hss}}  
\def \lta{\mathrel{\spose{\lower 3pt\hbox{$\sim$}}\raise  2.0pt\hbox{$<$}}}
\def \gta{\mathrel{\spose{\lower  3pt\hbox{$\sim$}}\raise 2.0pt\hbox{$>$}}}

\def \kms {\ifmmode  \,\rm km\,s^{-1} \else $\,\rm km\,s^{-1}  $ \fi }
\def \kpc {\ifmmode  {\rm~kpc}  \else ${\rm~kpc}$\fi}  
\def \pc {\ifmmode  {\rm~pc}  \else ${\rm~pc}$ \fi  }  
\def \Gyr {\ifmmode  {\rm~Gyr}  \else ${\rm~Gyr}$\fi}
\def \Msun {\ifmmode M_{\odot} \else $M_{\odot}$ \fi} 
\def \Lsun {\ifmmode L_{\odot} \else $L_{\odot}$ \fi} 
\def \Rsun {\ifmmode R_{\odot} \else $R_{\odot}$ \fi} 
\def \Msunpyr {\ifmmode M_{\odot}{\rm~yr}^{-1} \else $M_{\odot}{\rm~yr}^{-1}$ \fi} 
\def \hMsun {\ifmmode h^{-1}\,\rm M_{\odot} \else $h^{-1}\,\rm M_{\odot}$ \fi}

\def \LCDM {\ifmmode \Lambda{\rm CDM} \else $\Lambda{\rm CDM}$ \fi}
\def \sig8 {\ifmmode \sigma_8 \else $\sigma_8$ \fi} 
\def \OmegaM {\ifmmode \Omega_{\rm M} \else $\Omega_{\rm M}$ \fi} 
\def \OmegaL {\ifmmode \Omega_{\rm \Lambda} \else $\Omega_{\rm \Lambda}$\fi} 
\def \Deltavir {\ifmmode \Delta_{\rm vir} \else $\Delta_{\rm vir}$ \fi}
\def \rhocrit {\ifmmode \rho_{\rm crit} \else $\rho_{\rm crit}$ \fi}
\def \rhou {\ifmmode \rho_{\rm u} \else $\rho_{\rm u}$ \fi}
\def \zc {\ifmmode z_{\rm c} \else $z_{\rm c}$ \fi}

\def \rhos {\ifmmode \rho_{\rm s} \else $\rho_{\rm s}$ \fi} 
\def \rs {\ifmmode r_{\rm s} \else $r_{\rm s}$ \fi} 
\def \cvir {\ifmmode c_{\rm vir} \else $c_{\rm vir}$ \fi} 
\def \Rvir {\ifmmode r_{\rm vir} \else $R_{\rm vir}$ \fi}
\def \Vvir {\ifmmode V_{\rm  vir} \else  $V_{\rm vir}$  \fi} 
\def \Mvir {\ifmmode M_{\rm  vir} \else $M_{\rm  vir}$ \fi}  
\def \Nvir {\ifmmode N_{\rm  vir} \else $N_{\rm  vir}$ \fi}  
\def \Jvir {\ifmmode J_{\rm vir} \else $J_{\rm vir}$ \fi} 
\def \Evir {\ifmmode E_{\rm vir} \else $E_{\rm vir}$ \fi} 
\def \vvir {\ifmmode v_{\rm vir} \else $v_{\rm vir}$ \fi} 
\def \lam {\ifmmode \lambda  \else $\lambda$ \fi} 
\def \lamp {\ifmmode \lambda^{\prime} \else $\lambda^{\prime}$  \fi} 
\def \Vmax {\ifmmode V_{\rm  max} \else  $V_{\rm max}$  \fi} 
\def \Mdm {\ifmmode M_{\rm  dm} \else $M_{\rm  dm}$ \fi}

\def \Mgas {\ifmmode M_{\rm gas} \else $M_{\rm gas}$ \fi} 
\def \Mcg {\ifmmode M_{\rm cg} \else $M_{\rm cg}$\fi} 
\def \Mhg {\ifmmode M_{\rm hg} \else $M_{\rm hg}$ \fi} 
\def \Mdisc {\ifmmode M_{\rm disc} \else $M_{\rm disc}$ \fi} 
\def \Md {\ifmmode M_{\rm d} \else $M_{\rm d}$ \fi} 
\def \Mda {\ifmmode M_{\rm d,0\%} \else $M_{\rm d,0\%}$ \fi} 
\def \Mdb {\ifmmode M_{\rm d,20\%} \else $M_{\rm d,20\%}$ \fi} 
\def \Mdc {\ifmmode M_{\rm d,40\%} \else $M_{\rm d,40\%}$ \fi} 
\def \md {\ifmmode m_{\rm d} \else $m_{\rm d}$ \fi} 
\def \Mb {\ifmmode M_{\rm b} \else $M_{\rm b}$ \fi} 
\def \Mbh {\ifmmode M_{\rm b,pri} \else $M_{\rm b,pri}$ \fi} 
\def \Mbs {\ifmmode M_{\rm b,sat} \else $M_{\rm b,sat}$ \fi} 
\def \zo {\ifmmode z_{0} \else $z_{0}$ \fi} 
\def \rd {\ifmmode r_{\rm d} \else $r_{\rm d}$ \fi}
\def \rg {\ifmmode r_{\rm g} \else $r_{\rm g}$ \fi}
\def \rb {\ifmmode r_{\rm b} \else $r_{\rm b}$\fi}
\def \rs {\ifmmode r_{\rm s} \else $r_{\rm s}$\fi}
\def \rc {\ifmmode r_{\rm c} \else $r_{\rm c}$\fi}
\def \rvir {\ifmmode r_{\rm vir} \else $r_{\rm vir}$\fi}
\def \rbh {\ifmmode r_{\rm b,pri} \else $r_{\rm b,pri}$ \fi} 
\def \rbs {\ifmmode r_{\rm b,sat} \else $r_{\rm b,sat}$ \fi}

\title[Hot gaseous halo in major mergers] 
{The effects of a hot gaseous halo in galaxy major mergers}

\author[B. P. Moster et al.] {Benjamin P. Moster$^{1,2}$
 \thanks{moster@mpia.de}, Andrea V. Macci\`o$^{1}$, Rachel S. Somerville$^{3,4}$,
 \newauthor{Thorsten Naab$^{2}$, T. J. Cox$^{5}$}\\ 
  $^1$ Max-Planck-Institut f\"ur Astronomie, K\"onigstuhl 17, 69117 Heidelberg, Germany\\ 
  $^2$ Max-Planck Institut f\"ur Astrophysik, Karl-Schwarzschild Stra\ss e 1, 85748 Garching, Germany\\
  $^3$ Space Telescope Science Institute, Baltimore MD 21218\\
  $^4$ Department of Physics and Astronomy, Johns Hopkins University, Baltimore MD 21218\\
  $^5$ Carnegie Observatories, 813 Santa Barbara Street, Pasadena, CA 91101, USA\\
}

\begin{document} 
              
\date{\today}
              
\pagerange{\pageref{firstpage}--\pageref{lastpage}}\pubyear{2011} 
 
\maketitle 

\label{firstpage}
             
\begin{abstract}
Cosmological hydrodynamical simulations as well as observations
indicate that spiral galaxies are comprised of five different
components: dark matter halo, stellar disc, stellar bulge, gaseous
disc and gaseous halo.  While the first four components have been
extensively considered in numerical simulations of binary galaxy
mergers, the effect of a hot gaseous halo has usually been neglected
even though it can contain up to 80\% of the total gas within the
galaxy virial radius. We present a series of hydrodynamic simulations
of major mergers of disc galaxies, that for the first time include a diffuse,
rotating, hot gaseous halo. Through cooling and accretion, the hot halo can
dissipate and refuel the cold gas disc before and after a merger. This cold gas can
subsequently form stars, thus impacting the morphology and kinematics
of the remnant. Simulations of isolated systems with total mass
$M\sim10^{12}\Msun$ show a nearly constant star formation rate of
$\sim5\Msunpyr$ if the hot gaseous halo is included, while the star
formation rate declines exponentially if it is neglected.
We conduct a detailed study of the star formation efficiency during
mergers and find that the presence of a hot gaseous halo reduces the
starburst efficiency ($e=0.5$) compared to 
simulations without a hot halo ($e=0.68$).
The ratio of the peak star formation rate in mergers compared to isolated
galaxies is reduced by almost an order of magnitude (from 30 to 5).
Moreover we find cases where the stellar mass of the merger remnant is
lower than the sum of the stellar mass of the two progenitor galaxies
when evolved in isolation. This suggests a revision to semi-analytic
galaxy formation models which assume that a merger always leads to
enhanced star formation.
In addition, the bulge-to-total ratio after a major
merger is decreased if hot gas is included in the halo, due to the
formation of a more massive stellar disc in the remnant.
We show that adding the hot gas component has a significant effect on
the kinematics and internal structure of the merger remnants, like an increased
abundance of fast rotators and an $r^{1/4}$ surface brightness profile at small scales.
The consequences on the population of elliptical galaxies formed by disc
mergers are discussed.

\end{abstract}


\setcounter{footnote}{1}

%%%%%%%%%%%%%%%%%%%%%%%%%%%%%%%%%%%%%%%%%%%%%%%%%%%
%%%%%%%%%%%%%%%%%%%
%% SECTION 1: INTRODUCTION
%%%%%%%%%%%%%%%%%%%%%%%%%%%%%%%%%%%%%%%%%%%%%%%%%%%
%%%%%%%%%%%%%%%%%%%
\section{Introduction}
\label{sec:intro}

Galaxy mergers, both major (near-equal mass) and minor (unequal mass),
are a generic prediction of structure assembly in the Cold Dark Matter
(CDM) model \citep[e.g.][]{white1978,davis1985}. Mergers are an
important component of the modern theory of galaxy formation and are
thought to be a significant physical driver of many galaxy properties.
Major mergers can transform disc-dominated galaxies into spheroidals
\citep{toomre1977,negroponte1983,hernquist1992,naab2003,cox2006b},
enhance star formation
\citep{larson1978,kennicutt1987,barnes1991,mihos1994,mihos1996,barton2003,
  cox2008} and possibly trigger Active Galactic Nuclei (AGN) activity
\citep{sanders1988,hernquist1989,springel2005b,hopkins2006,johansson2009,
younger2009,debuhr2010,debuhr2011}.
Minor mergers can thicken the galactic disc
\citep{quinn1993,velazquez1999,brook2004,bournaud2005,kazantzidis2008,
  villalobos2008,read2008,purcell2009,moster2010b} and create a diffuse stellar halo
\citep{murante2004,bullock2005,bell2008,murante2010}.  Observations
find large samples of recent merger remnants in the local universe
\citep{schweizer1982,lake1986,doyon1994,
  shier1998,genzel2001,dasyra2006} and that $\sim5-10\%$ of low- and
intermediate-mass galaxies are in the process of merging \citep{bridge2010}.
An average $\sim M_*$ galaxy has experienced about one major merger since
$z\sim2-3$ \citep{bridge2007,kartaltepe2007,conselice2008,lin2008,
lotz2008,conselice2009,jogee2009}.

Since the ``merger hypothesis'' was proposed by \citet{toomre1972}
there have been many publications on numerical simulations studying
the dynamics of galaxy mergers and the properties of merger remnants
\citep{gerhard1981,negroponte1983,barnes1988,
  hernquist1993,barnes1996,dimatteo2005,naab2006b,cox2006,jesseit2007,hoffman2009}.
In recent years several groups have produced very large libraries of
major merger simulations, considering different orbital parameters of
the merger, the effect of the implemented gas physics, and different
star formation algorithms and feedback schemes
\citep{robertson2006,cox2006b, naab2006,bois2010}. One clear outcome
of those simulations has been the need to include gas physics in
order to reproduce basic properties of observed galaxies. Gas plays an
important role in shaping the properties of galaxy mergers and their
remnants, because unlike stars and dark matter, gas can cool
radiatively and therefore lose kinetic energy efficiently.
Furthermore, gas can lose angular momentum due to gravitational
torques during merger events.  After the merger, an extended cold gas
disc can form in the remnant from high angular momentum gas in the
outer regions of the progenitors. By the accumulation of cold gas in
the centre of a merger remnant the central potential changes
impacting the orbital configuration of the stars \citep{barnes1996}. Additionally,
the gas is eventually transformed into stars, leading to a new stellar
component which is rotating and kinematically cold. It has been shown
that in numerical simulations even major mergers of disc galaxies can
produce remnants with large discs if the cold gas fractions in the
progenitors are sufficiently high \citep{barnes2002,springel2005c,
  robertson2006b,naab2006,governato2007,governato2009}. This has also
been seen in observations of high-redshift discs that are probably
reforming after mergers
\citep{vandokkum2004,kassin2007,genzel2008,puech2008}.

Detailed studies of elliptical galaxies indicate that they can be
classified into two groups with respect to their structural properties
\citep{davies1983,bender1988a,bender1990,kormendy1996}.
Low and intermediate mass giant elliptical galaxies exhibit significant
rotation along the photometric major axis \citep[see e.g.][]{
emsellem2007,emsellem2011}. Although their rotation is
broadly consistent with models of isotropic rotators
it has been demonstrated by \citet{cappellari2007} and \citet{emsellem2011}
that rotating ellipticals are intrinsically more anisotropic than slowly rotating
massive ellipticals. This finding is in agreement with disc merger simulations
\citep{burkert2008}. The isophotes of rotating ellipticals tend to be discy. They have
power-law surface brightness profiles \citep{lauer1995,faber1997} and
show little or no radio and X-ray emission \citep{bender1989}. Large,
luminous spheroids, on the other hand, have box-shaped isophotes, show
flat cores and strong radio and X-ray emission. They rotate slowly,
exhibit kinematically decoupled components, and have a large
amount of minor-axis rotation \citep[see][for recent results from the
ATLAS$^{\rm 3D}$ survey]{cappellari2011,emsellem2011,krajnovic2011}.

The different physical properties of elliptical galaxies lead to the
viewpoint that the two classes have different formation histories and
form through different mechanisms. Many studies find that
collisionless merger simulations lead to slowly rotating, pressure
supported, anisotropic remnants
\citep{negroponte1983,barnes1988,hernquist1992}. Along this line of
reasoning, \citet{naab1999}, \citet{bendo2000} and \citet{naab2003}
argue that fast rotating low luminosity ellipticals are produced by
dissipationless minor mergers. However, remnants of dissipationless
disc galaxies are in conflict with observations, as they do not have
de Vaucouleurs profiles in the inner parts. The reason for this is
that low phase-space density spirals cannot produce the high central
phase-space densities of ellipticals --- according to Liouville's
theorem, phase-space density is conserved during a collisionless
process \citep{carlberg1986}. This problem could only be overcome in
dissipationless simulations by including large bulge components in the
progenitor systems \citep{hernquist1993b,naab2006b}.

Another way to circumvent this problem is to take into account the gas
component in the progenitors, which can increase the phase-space
density through radiation \citep{lake1989}. It has been argued by
\citet{kormendy1996} and \citet{faber1997} that the observed stellar
discs embedded in rotating ellipticals
\citep{rix1990,ferrarese1994,scorza1998,rix1999,lauer2005,krajnovic2008}
experienced dissipation during their formation while non-rotating
systems would form from pure dissipationless mergers. This idea has
been explored by several studies
\citep{bekki1997,bekki1998,naab2006,jesseit2009,hopkins2008,
hopkins2009a,hopkins2009b,hopkins2009c}. Focusing on equal
mass mergers, \citet{cox2006} confirmed that slowly rotating
anisotropic spheroids can be formed in dissipationless
simulations. Furthermore they show that if a massive gaseous disc is
included in the progenitors, a consistent fraction of merger orbits
leads to systems with significant rotation. Those remnants are able to
reproduce the observed distribution of projected ellipticities,
rotation parameters and isophotal shapes. One open issue is that these
results rely on a relatively high (possibly too high) gas content in
the discs of the merger progenitors and that only a fraction of the
considered orbits lead to fast rotators. It remains unclear whether
the abundance of disc galaxies with such gas fractions is high enough
in order to explain the large number of fast rotating elliptical
galaxies observed in the local universe
\citep{zeeuw2002,emsellem2007,cappellari2007,emsellem2011}.

The initial conditions for almost all hydrodynamic simulations of galaxy mergers
performed so far have considered only cold gas that is present within the progenitor
discs at the start of the simulation. However, semi-analytic models of galaxy formation
\citep{kauffmann1993,bower2006,delucia2007,monaco2007,somerville2008a}
as well as full cosmological hydrodynamic simulations
\citep[e.g.][]{toft2002,sommerlarsen2006,johansson2009b,rasmussen2009,
  stinson2010,hansen2010} both predict a large amount of hot gas in
quasi-hydrostatic equilibrium within the gravitational potential of
the dark matter halo. Smooth accretion of gas from these haloes can
grow the discs of spiral galaxies \citep{abadi2003,sommerlarsen2003,guo2008}.
As the gas mainly cools via thermal bremsstrahlung it radiates primarily in the soft
X-ray band.

Since this halo X-ray luminosity depends on the mass of the system,
observational studies are naturally biased towards very massive
haloes. Hence, hot gas in massive ellipticals has been observed in
galaxy groups and clusters and in isolated systems \citep[see] [and
  references therein]{mathews2003}. In smaller haloes, however, the
gas temperature can be lower than $10^6$K and thus difficult to
observe with current X-ray telescopes, as the radiation is only
detectable in very soft bands. Still, studies performed in the last
few years have identified X-ray emission from diffuse hot gas in
various energy bands \citep{benson2000,wang2003,strickland2004,
  wang2005,tuellmann2006,li2007,sun2009,owen2009}. In the MW, X-ray
absorption lines produced by local hot gas have been detected in the
spectra of several AGN. \citet{sembach2003} and \citet{tripp2003} have
argued that this emission comes from the interface between warm clouds
and the ambient hot medium. Given the existence of hot gaseous haloes
around galaxies, it is important to include these in simulations of
isolated galaxies and mergers.

In the last few years there have been several studies that have taken
a gaseous halo into account. \citet{mastropietro2005} and
\citet{mastropietro2009} employ a hot halo component to study the
hydrodynamic and gravitational interaction between the Large
Magellanic Cloud and the MW. \citet{sinha2009} ran merger simulations
of galaxies consisting of dark matter and hot halo gas in order to
study the change of temperature and X-ray luminosity induced by shocks
during the mergers. Those studies, however, use adiabatic simulations,
neglecting cooling and star formation. Focusing on isolated systems,
\citet{kaufmann2006} and \citet{kaufmann2009} simulated systems
consisting of a dark matter and a hot halo component to study the
evolution and cooling behaviour of the hot halo and the formation of
discs via cooling flows. Similarly, \citet{viola2008} ran simulations
of a gas halo embedded in a dark matter halo to study the cooling
process and assess how well simple models can represent it. So far, no
study has included a rotating, cooling gaseous halo in merger
simulations.

The goal of this paper is to test the impact of including such a large gas
reservoir on star formation and remnant properties in major mergers of disc galaxies.
For the first time we include a gradually cooling hot gas halo (as
expected in a cosmological context) in our galaxy models. We
present a detailed analysis of the effects of this new gas
component and focus our attention on three main quantities: the star
formation efficiency during and after the merger, and the morphology
and kinematics of merger remnants. 

The paper is organized as it follows: in section \ref{sec:nsim} we
provide a brief summary of the {\sc GADGET-2} code and the initial
conditions. We also explain how the initial conditions have been
set up to include a rotating hot gaseous halo, and show the results
of simulations constraining the initial spin of
this hot halo. In section \ref{sec:maj} we present our main results
for the major merger simulations, focusing on the differences between
the simulations with and without a hot gaseous component. We present
star formation rates, starburst efficiencies and structural and
kinematic properties of the remnants. Finally, in section
\ref{sec:conc} we summarize and discuss our results and compare them
to previous studies that have neglected the gaseous halo.

%%%%%%%%%%%%%%%%%%%%%%%%%%%%%%%%%%%%%%%%%%%%%%%%%%%
%%%%%%%%%%%%%%%%%%%
%% SECTION 2: CODE AND INITIAL CONDITIONS
%%%%%%%%%%%%%%%%%%%%%%%%%%%%%%%%%%%%%%%%%%%%%%%%%%%
%%%%%%%%%%%%%%%%%%%
\section{Numerical Simulations} 
\label{sec:nsim}

\subsection{Numerical Code} 
\label{sec:ncode}

All numerical simulations in this work were performed using the
parallel TreeSPH-code {\sc GADGET-2} \citep{springel2005a}. Gas
dynamics is followed with the Lagrangian smoothed particle
hydrodynamics \citep[SPH][]{lucy1977,gingold1977,monaghan1992,springel2010}
technique, employed in a formulation that manifestly conserves energy
and entropy \citep{springel2002}. Radiative cooling is implemented for
a primordial mixture of hydrogen and helium following \citet{katz1996},
with a spatially uniform time-independent local photo-ionizing UV
background in the optically thin limit \citep{Haardt1996}.

\begin{table*}
 \centering
 \begin{minipage}{140mm}
  \caption{Parameters kept constant for all simulations. Masses are in units of  $10^{10}\Msun$, scale and 
softening lengths are in units of \kpc~and $\pc$, respectively}
  \begin{tabular}{@{}lrrrrrrrrrrrr@{}}
  \hline
  System & \Mdm & \Mb & \rd & \rg & \rb & \zo & $c$ & $\lambda$ & $N_{\rm dm}$ & $N_{\rm bulge}$ & $\epsilon_{\rm DM}$ & $\epsilon_{\rm stars}$\\
 \hline
 \hline
Z1 & 80 & 0.500 & 2.50 & 3.75 & 0.50 & 0.6 & 3.64 & 0.030 & 500 000 & 20 000 & 400 & 100\\
G3 & 110 & 0.890 & 2.85 & 8.55 & 0.62 & 0.4 & 6.00 & 0.050 & 240 000 & 20 000 & 400 & 100\\
\hline
\label{t:conpar}
\end{tabular}
\end{minipage}
\end{table*}

For modelling star formation and the associated heating by supernovae
(SN) we follow the sub-resolution multiphase ISM model developed by
\citet{springel2003}. In this model, a thermal instability is assumed
to operate above a critical density threshold $\rho_{th}$, producing a
two-phase medium which consists of cold clouds embedded in a tenuous
gas at pressure equilibrium. Stars are formed from the cold clouds on
a timescale chosen to match observations \citep{kennicutt1998} and
short-lived stars supply an energy of $10^{51}$ ergs to the
surrounding gas as supernovae. This energy heats the diffuse phase of
the ISM and evaporates cold clouds, such that a self-regulated cycle
for star formation is established.  The threshold density $\rho_{th}$
is determined self-consistently by demanding that the equation of
state (EOS) is continuous at the onset of star formation. In a subset
of simulations we include SN-driven galactic winds as proposed by
\citet{springel2003}. In this model the mass-loss rate carried by the
wind is proportional to the star formation rate (SFR) $\dot M_w= \eta
\dot M_*$, where the mass-loading-factor $\eta$ quantifies the wind
efficiency. Furthermore, the wind is assumed to carry a fixed fraction
of the supernova energy, such that there is a constant initial wind
speed $v_w$ (energy-driven wind).

We adopt the standard parameters for the multiphase model in order to
match the Kennicutt Law as specified in \citet{springel2003}. The star
formation timescale is set to $t_*^0 = 2.1\Gyr$, the cloud evaporation
parameter to $A_0 = 1000$ and the SN ``temperature'' to $T_{\rm
  SN}=10^8{\rm~K}$.  We employ a \citet{salpeter1955} initial mass
function (IMF) which sets the mass fraction of massive stars
$\beta=0.1$. For the galactic winds we adopt a mass-loading factor of
$\eta = 2$ which is motivated by observations
\citep{martin1999,rupke2005,martin2005}, and a wind speed of $v_w \sim
480 \kms$, typical for a Milky Way (MW)-like galaxy at low redshift. We
do not include feedback from accreting black holes (AGN feedback) in
our simulations.

\subsection{Galaxy Models} 
\label{sec:nmod}

To construct the galaxy models used in our simulations we apply and
extend the method described in \citet{springel2005b}. Each system is
composed of a cold gaseous disc, a stellar disc and a stellar bulge
with masses \Mcg, \Mdisc and \Mb embedded in a halo that consists of
hot gas and dark matter with masses \Mhg and \Mdm.

The gaseous and stellar discs are rotationally supported and have
exponential surface density profiles.  The scale length of the gaseous
disc \rg is related to that of the stellar disc \rd by $\rg = \chi
\rd$. The vertical structure of the stellar disc is described by a
radially independent sech$^2$ profile with a scale height $z_0$, and
the vertical velocity dispersion is set equal to the radial velocity
dispersion. The gas temperature is fixed by the EOS, rather than the
velocity dispersion. The vertical structure of the gaseous disc is
computed self-consistently as a function of the surface density by
requiring a balance of the galactic potential and the pressure given
by the EOS. The spherical stellar bulge is non-rotating and is
constructed using the \citet{hernquist1990} profile with a scale
length \rb. The dark matter halo has a \citet{hernquist1990} profile
with a scale length \rs, a concentration parameter $c=\rvir/\rs$ and a
halo spin $\lambda$.

\subsubsection{Modelling the hot gaseous halo}
For MW-like galaxies there are no existing observations that constrain
the profile of the gaseous hot halo. Therefore, we choose the
observationally motivated $\beta$-profile
\citep{cavaliere1976,jones1984,eke1998} which is commonly used to
describe hot gas in galaxy clusters:
\begin{equation}
\rho_{\rm hg}(r) = \rho_0 \left[1+\left(\frac{r}{\rc}\right)^2\right]^{-\frac{3}{2}\beta}\;.
\end{equation}
It has three free parameters: the central density $\rho_0$, the core
radius \rc~and the outer slope parameter $\beta$.  The temperature
profile can be fixed by assuming isotropy and hydrostatic equilibrium
inside the galactic potential. The halo temperature at a given radius
$r$ is then determined by the cumulative mass distribution $M(r)$ of
the dark matter, stellar and gaseous components beyond $r$ and by the
density profile $\rho_{\rm hg}(r)$ of the hot gas:
\begin{equation}
T(r) = \frac{\mu m_p}{k_B} \frac{1}{\rho_{\rm hg}(r)} \int_r^\infty \rho_{\rm hg}(r) \frac{G M(r)}{r^2} dr\; ,
\end{equation}
where $m_p$ is the proton mass, $G$ and $k_B$ are the gravitational
and Boltzmann constants and $\mu$ is the mean molecular weight.

In addition the hot gaseous halo is rotating around the spin axis of
the disc. The angular momentum of the hot gaseous halo is set by
requiring that the specific angular momentum of the the gas $j_{\rm
  hg} = J_{\rm hg} / M_{\rm hg} $ is a multiple of the specific
angular momentum of the dark matter halo $j_{\rm dm} = J_{\rm dm} /
M_{\rm dm} $ such that $ j_{\rm hg} = \alpha j_{\rm dm}$. A value of
$\alpha=1$ matches the commonly adopted assumption that there is no
angular momentum transport between the spherical dark matter halo and
the gaseous halo. The angular momentum distribution is then assumed to
scale with the product of the cylindrical distance from the spin axis
$R$ and the circular velocity at this distance: $j(R) \propto R \;
v_{\rm circ}(R)$. The vertical velocity of the gas halo particles is
set equal to zero.

\subsubsection{Parameters of the gaseous halo} 
\label{sec:amgh}

\begin{table*}
 \centering
 \begin{minipage}{140mm}
  \caption{Parameters for the different simulation runs. Masses are in
    units of $10^{10}\Msun$, and softening lengths are in units of $\pc$.}
  \begin{tabular}{@{}lrrrrrrrrrrr@{}}
  \hline
  Run & $f_{\rm gas}$ & \Mdisc & \Mcg & \Mhg & $\alpha$ &  $N_{\rm disc}$ & $N_{\rm cg}$ & $N_{\rm hg}$ & $\epsilon_{\rm gas}$ & $\theta$\\
 \hline
 \hline
Z1 & 0.40 & 2.00 & 1.33 & 0.0 & - & 100 000 & 33 333 & 0 & 140 & -\\
Z1A1 & 0.40 & 2.00 & 1.33 & 12.0 & 1 & 100 000 & 33 333 & 375 000 & 140 & -\\
Z1A2 & 0.40 & 2.00 & 1.33 & 12.0 & 2 & 100 000 & 33 333 & 375 000 & 140 & -\\
Z1A4 & 0.40 & 2.00 & 1.33 & 12.0 & 4 & 100 000 & 33 333 & 375 000 & 140 & -\\
Z1A8 & 0.40 & 2.00 & 1.33 & 12.0 & 8 & 100 000 & 33 333 & 375 000 & 140 & -\\
\hline
G3 & 0.23 & 4.11 & 1.20 & 0.0 & - & 100 000 & 15 000 & 0 & 140 & $30^{\circ}$\\
G3f0 & 0.00 & 5.31 & 0.00 & 0.0 & - & 130 000 & 0 & 0 & - & $30^{\circ}$\\
G3f4 & 0.40 & 3.20 & 2.11 & 0.0 & - & 80 000 & 25 000 & 0 & 140 & $30^{\circ}$\\
G3f8 & 0.80 & 1.07 & 4.24 & 0.0 & - & 30 000 & 50 000 & 0 & 140 & $30^{\circ}$\\
G3h & 0.23 & 4.11 & 1.20 & 11.0 & 4 & 100 000 & 15 000 & 175 000 & 140 & $30^{\circ}$\\
G3hf0 & 0.00 & 5.31 & 0.00 & 11.0 & 4 & 130 000 & 0 & 175 000 & 140 & $30^{\circ}$\\
G3hf4 & 0.40 & 3.20 & 2.11 & 11.0 & 4 & 80 000 & 25 000 & 175 000 & 140 & $30^{\circ}$\\
G3hf8 & 0.80 & 1.07 & 4.24 & 11.0 & 4 & 30 000 & 50 000 & 175 000 & 140 & $30^{\circ}$\\
G3hX & 0.23 & 4.11 & 1.20 & 5.5 & 4 & 100 000 & 15 000 & 87 500 & 140 & $30^{\circ}$\\
\hline
\label{t:varpar}
\end{tabular}
\end{minipage}
\end{table*}

The hot gaseous halo is then described by four parameters: the central
density $\rho_0$, the core radius \rc, the slope parameter $\beta$ and
the spin factor $\alpha$.  For the density profile, we adopt $\beta =
2/3$ \citep{jones1984} and $\rc=0.22\rs$ \citep{makino1998}.  Thus
there is a central core with a constant density and the slope in the
outer halo is $-2$.

In order to demonstrate the maximum effect that the hot gaseous halo
can have, we fix $\rho_0$ such that the baryonic fraction within
\rvir~(stellar and gaseous discs, bulge and gaseous halo) is the
universal value, as this is the maximum fraction that a gaseous halo can
achieve. We also test an intermediate case, where we adopt only half
of this maximal hot gas mass to investigate whether the results fall
between the extreme cases (no and maximal gaseous halo). The results
of this test are presented in section \ref{sec:halfhalo}. Note,
however, that even for our maximum hot gaseous halo mass we do not
violate constraints from X-ray observations; see section
\ref{sec:conc} for a more detailed discussion of this point.

This leaves one key free parameter which needs to be constrained, the
spin factor $\alpha$.  In high resolution cosmological simulations
one finds that at low redshift ($z\lta2$) $\alpha$ is generally larger
than 1, as feedback processes preferentially remove low angular
momentum material from the halo \citep{governato2010}.

In order to fix the spin factor for MW-like galaxies we model a
typical MW-like galaxy at $z=1$ and let it evolve to $z=0$ using
different values for $\alpha$. We then employ two observational
constraints (stellar mass and disc size) to determine the correct
$\alpha$. A typical MW-like galaxy with $M_{\rm dm}(z=0) \sim
10^{12}\Msun$ has a halo concentration of $c=8.43$ at $z=0$
\citep{maccio2008} which fixes both parameters of the Hernquist
profile. Assuming that the halo profile does not change from $z=1$ to
the present, we can compute the virial mass and the halo concentration
given the higher background density at $z=1$, resulting in $M_{\rm dm,
  z=1} = 8.0\times10^{11}\Msun$ and $c_{z=1} = 3.64$. Using the
redshift-dependent stellar-to-halo mass relation derived by
\citet{moster2010a}, converted to a Salpeter IMF, we assign a stellar
mass of $M_{*,z=1}=2.5\times10^{10}\Msun$ to the system.  Distributing
80\% of this stellar mass into the exponential disc yields a stellar
disc mass of $\Mdisc=2.0 \times10^{10}\Msun$ and a bulge mass of
$\Mb=5\times10^{9}\Msun$. A bulge scale length of $\rb=0.5\kpc$ is
assumed. In order to determine the mass of the gaseous disc we use the
recipe by \citet{stewart2009} which is based on data by
\citet{mcgaugh2005} and \citet{erb2006}. This recipe predicts that
central galaxies with $M_*=2.5\times10^{10}\Msun$ at $z=1$ have a gas
fraction of $f_{\rm gas}= \Mcg / (\Mcg + \Mdisc) = 0.4$ which yields a
mass for the gaseous disc of $\Mcg =1.33\times10^{10}\Msun$. The scale
length for the stellar disc at $z=1$ is set to $\rd=2.4\kpc$,
consistent with observations from GEMS
\citep{barden2005,somerville2008b}, and corresponds to a spin
parameter of $\lambda = 0.03$ assuming the scalings of
\citet{mo1998}. We assume that the scale length of the
gaseous disc $\rg$ is a factor of $\chi=1.5$ larger. The scale height
of the stellar disc is set to $\zo = 0.6\kpc$, typical for MW-like
galaxies \citep{schwarzkopf2000,yoachim2006}. The mass of the hot
gaseous halo within \rvir~is $\Mhg = 1.2 \times10^{11}\Msun$, such
that the baryonic fraction within \rvir~is the universal value. The
system is modeled with $N_{\rm dm}=500\textrm{,}000$ dark matter,
$N_{\rm disc}=100\textrm{,}000$ stellar disc, $N_{\rm
  cg}=33\textrm{,}333$ gaseous disc, $N_{\rm bulge}=20\textrm{,}000$
bulge and $N_{\rm hg}= 375\textrm{,}000$ gaseous halo particles. We
set the gravitational softening length to $\epsilon = 100\pc, 140\pc$
and $400\pc$ for stellar, gas and dark matter particles,
respectively. We summarize the parameters that are kept constant for
all simulations in Table~\ref{t:conpar}, and parameters that differ
for the various simulation runs are summarized in Table
\ref{t:varpar}.

In order to fix the angular momentum of the gaseous halo, we run five
simulations with different values of the spin factor $\alpha$. These
include: one simulation with no gaseous halo at all (Z1), and
simulations with $\alpha = 1, 2, 4$ and $8$ (Z1A1, Z1A2, Z1A4,
Z1A8). We evolve these simulations in isolation from $z=1$ to $z=0$
which corresponds to $7.6\Gyr$. The simulations include stellar winds.
For every time-step of the simulation we measure the virial mass of
the dark matter halo (given the background density at this epoch) and
use the empirical stellar-to-halo mass constraints derived by
\citet{moster2010a} at that redshift to compute the mean stellar mass
expected for this halo mass. We compare this to the stellar mass
measured in the simulation. We also compute the radial density profile
of the stellar disc and measure the exponential scale length.
These values are compared to the mean observed
scale length for disc galaxies of corresponding stellar mass at the
given redshift \citep{barden2005,somerville2008b}. For example, the
considered halo has a virial mass of $\Mdm=0.9\times10^{12}\Msun$ at
$z=0.4$. The mean stellar mass of galaxies in a halo of this mass at
this redshift is $M_*=4\times10^{10}\Msun$ and the mean observed scale
length for galaxies of this stellar mass is $\rd=2.82\kpc$. We require
that the stellar mass and scale length measured in the simulations
should obey these constraints.

\begin{figure}
\psfig{figure=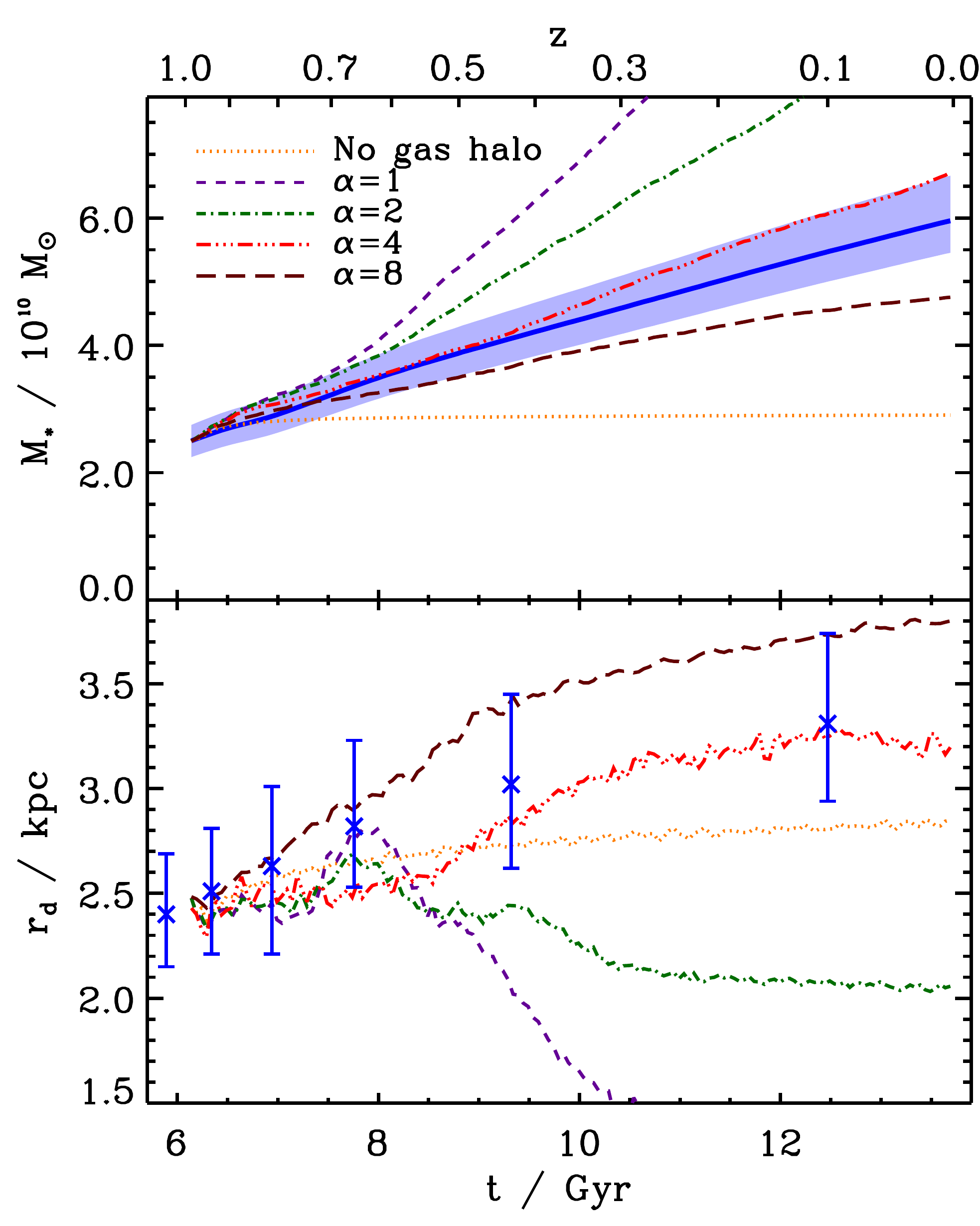,width=0.47\textwidth}
\caption{\textit{Upper panel:} Stellar mass as a function of time for
  different values of the spin parameter for the hot gas halo
  $\alpha$. The values measured in the simulations (lines) are compared
  to the observational constraints from \citet[][bold line and shaded
  area]{moster2010a}.
  \textit{Lower panel:} Disc scale length as function of time
  for different $\alpha$. Symbols show the observed values by
  \citet{barden2005}. Only the simulation with $\alpha=4$ agrees well
  with both constraints.}
\label{fig:spin}
\end{figure}

We show the results of the simulations in Figure~\ref{fig:spin}. For
the simulation without a gaseous halo the evolution of the scale
length agrees well with the observed values at $z\gta0.4$, while after
that it does not grow rapidly enough and ends up being too small. The
stellar mass evolution, however, completely disagrees with that of a
typical galaxy of this halo mass. The reason for this is the lack of
cold gas from which stars can form.  As the initial cold gaseous disc
is depleted quickly, but does not get refuelled from an external
reservoir, there is not enough gas available to increase the disc mass
by a factor of two by $z=0$. This shows that the hot gaseous halo is
very important in order to reproduce the SF history of spiral galaxies
like our MW, by providing an extended supply of cold gas that can fuel
ongoing star formation.

The simulations including a gaseous halo show a much larger SFR, with the
value depending on the initial spin factor $\alpha$. For low values of
$\alpha=1$ and 2, the stellar mass of the galaxy increases too fast
and becomes too large at $z=0$, while the scale length quickly
decreases and reaches values that are much smaller than the observed
ones. These effects are due to the low angular momentum of the halo
gas at the start of the simulation. When the gas cools, it is no
longer supported by pressure, but only by angular momentum.  This
means that a lower initial angular momentum results in a lower orbital
distance and thus in a lower scale length. Additionally, since the
centrifugal barrier is lower, much more material can reach densities
where SF is active which yields a much higher SFR and thus a larger
stellar mass. For values of $\alpha \lta 3$, too much material
concentrates at small distances from the galactic centre which results
in overly massive and too compact discs compared to observations. On
the other hand, large values of $\alpha=8$ yield a stellar disc which
is not massive enough and which has a scale length that is too
large. Due to the larger initial angular momentum the cooling gas can
retain large orbital distances. This also leads to much less dense
star-forming material and thus to lower stellar masses. For values of
$\alpha \gta 6$, we find that gas settles in discs that are too large
and does not form enough stars compared to observations.

The model that agrees best with the observational constraints is the
one with a spin factor of $\alpha=4$. For this value we find a stellar
mass and a scale length which are consistent with the observational
constraints for MW-like galaxies. As a result we use this value
throughout the rest of this paper for all systems, which are all in
the halo mass range of the Milky-Way. We note that different values of
$\alpha$ (representing different merger histories, feedback
efficiencies, etc.)  can easily result in a range of stellar masses
and scale length, in agreement with the spread of these quantities for
observed spiral galaxies.

\subsection{Merger Simulation Parameters} 
\label{sec:npar}

To study equal mass mergers of disc galaxies we adopt the galaxy model
G3 used by \citet{cox2008}, which is tuned to match SDSS observations
of local spirals. The stellar mass of this galaxy was chosen to be
$M_*=5\times10^{10}\Msun$. The bulge-to-disc ratio of $B/D=0.22$ is
taken from \citet{dejong1996}, resulting in a stellar disc mass of
$\Mdisc=4.11\times10^{10}\Msun$ and a bulge mass of $\Mb=8.9\times
10^{9}\Msun$. The gas fraction in the disc of $0.23$ is motivated by
the gas-to-stellar mass scaling relation from \citet{bell2003} and
yields a mass for the cold gaseous disc of
$\Mcg=1.2\times10^{10}\Msun$.  The sizes of the stellar disc and bulge
were chosen to agree with the stellar mass-size relation of
\citet{shen2003} resulting in $\rd=2.85\kpc$ and $\rb=0.62\kpc$. The
scale length of the gaseous disc was assumed to be a factor of
$\chi=3$ larger than that of the stellar disc, such that
$\rg=8.55$. The scale height of the stellar disc was set to
$\zo=0.4\kpc$. The mass and concentration of the dark matter halo were
selected such that the rotation curves lie on the baryonic
Tully-Fisher relation \citep{bell2001,bell2003} which yields a virial
mass of $M_{\rm dm} = 1.1\times10^{12}\Msun$ and a concentration
parameter of $c=6$.  In order to explore the effects of different
initial progenitor gas fractions on starbursts and remnant
morphologies we construct additional galaxy models with modified disc
gas fractions of 0\%, 40\% and 80\% (G3f0, G3f4 and G3f8).  In some
simulations we extend this model to also include a hot gaseous halo
(G3hf0, G3h, G3hf4, G3hf8) with a mass of $\Mhg =
1.1\times10^{11}\Msun$ within \rvir~and a hot gas spin factor of
$\alpha=4$.  In addition, we construct a system with only half of the
maximal hot gas mass, i.e. $\Mhg = 5.5\times10^{10}\Msun$ within
\rvir~(G3hX).  We set the gravitational softening length to $\epsilon
= 100\pc, 140\pc$ and $400\pc$ for stellar, gas and dark matter
particles, respectively.

We follow \citet{cox2008} and choose a nearly unbound elliptical orbit
with an eccentricity of $\epsilon_{\rm orbit}= 0.95$,
a pericentric distance of $r_{\rm min}=13.6\kpc$ and an initial
separation of $d_{\rm start}=250\kpc$\footnote{At this initial
  distance the two hot gaseous haloes are already penetrating each
  other. This, in principle, could reduce the strength of the shock at
  the boundaries of the haloes during the merger. We explicitly tested
  this by running a simulation with $d_{\rm start}=500\kpc$ and found
  no difference in the temperature and density profiles. This is not
  surprising because of the very low density of the hot gas in the
  outer regions ($r>125\kpc$).}.
The resulting interactions are fast and nearly radial, consistent with
merger orbits found in cosmological $N$-body simulations
\citep{benson2005,khochfar2006}. We employ a prograde orbit; the spin
axes of the first galaxy and the orbital plane are aligned while the
second galaxy is inclined by $\theta=30^{\circ}$ with respect to the
orbital plane. We evolve all simulations for 5\Gyr, such that the
remnant galaxies are relaxed and do not show any obvious signs of a
recent merger event.

%%%%%%%%%%%%%%%%%%%%%%%%%%%%%%%%%%%%%%%%%%%%%%%%%%%
%%%%%%%%%%%%%%%%%%%%%
%% SECTION 3: MAJOR MERGERS
%%%%%%%%%%%%%%%%%%%%%%%%%%%%%%%%%%%%%%%%%%%%%%%%%%%
%%%%%%%%%%%%%%%%%%%%%
\section{Major mergers}
\label{sec:maj}

\begin{figure*}
\includegraphics[width=0.8\textwidth]{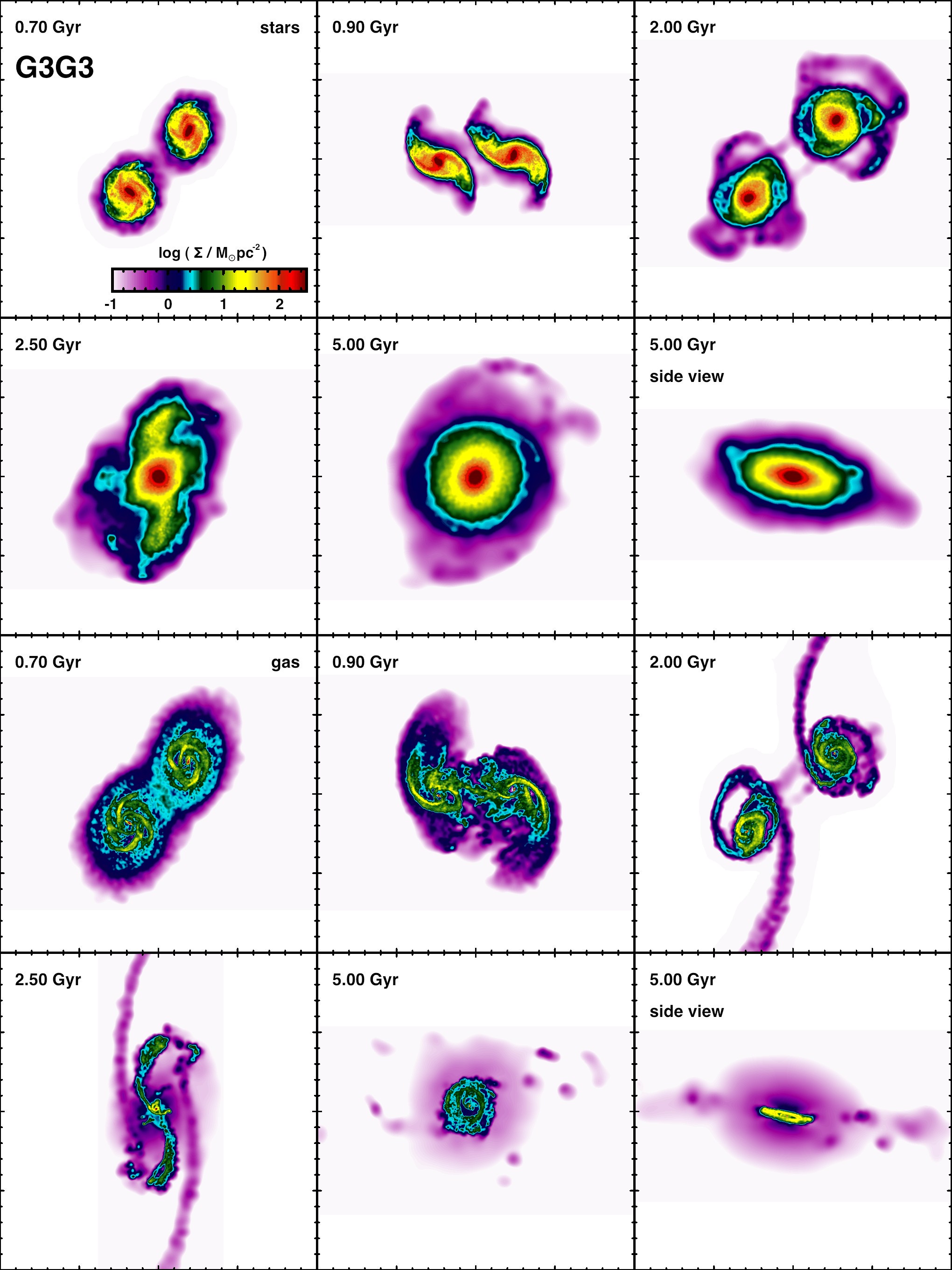}
\caption{Projected surface density for the stellar component (upper
  two rows) and the gaseous component (lower two rows) for the
  merger without hot gaseous haloes and without stellar winds as
  viewed in the orbital plane. Each panel measures 200\kpc~on a side
  and the time in\Gyr~is displayed in the upper left corner of each
  panel. The right-hand panels in the second and bottom columns show a
  side view of the final merger remnant. The images were created
  with {\sc splash} \citep{price2007}.}
\label{fig:G3G3}
\end{figure*}

\begin{figure*}
\includegraphics[width=0.8\textwidth]{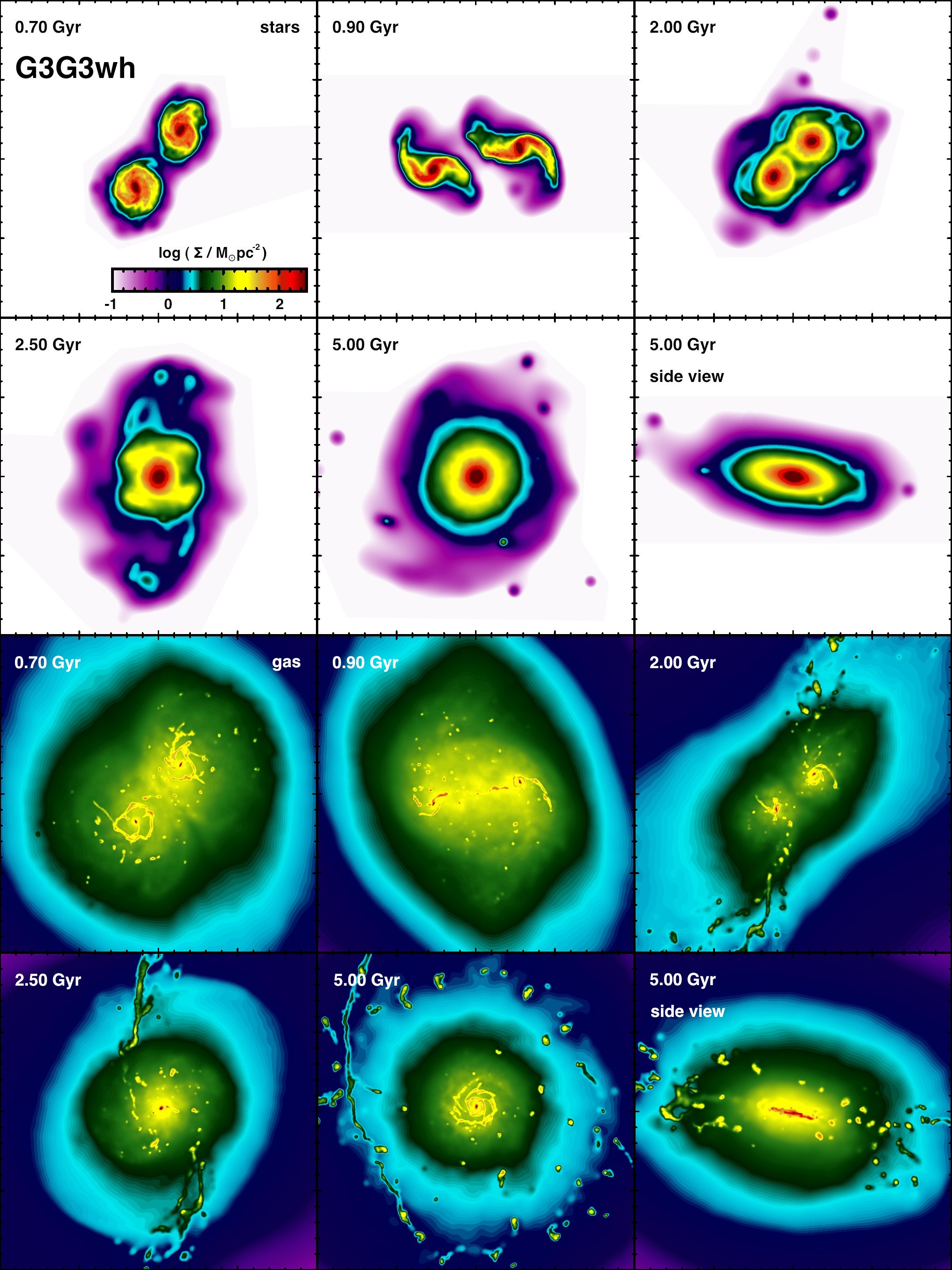}
\caption{Similar to Figure \ref{fig:G3G3}, but for the simulation with
  stellar winds and hot gaseous haloes.}
\label{fig:G3G3wh}
\end{figure*}

We have simulated the fiducial major merger using four different sets
of ingredients: one without stellar winds and without hot gaseous halo
(G3G3), one with stellar winds and no gaseous halo (G3G3w), one
without winds but with a gaseous halo (G3G3h) and finally one with
winds and a gaseous halo (G3G3wh). 
The surface density in the orbital plane is shown in Figures
\ref{fig:G3G3} (G3G3) and \ref{fig:G3G3wh} (G3G3wh) for the stellar
component (upper panels) and the gaseous component (lower panels).

\begin{figure*}
\psfig{figure=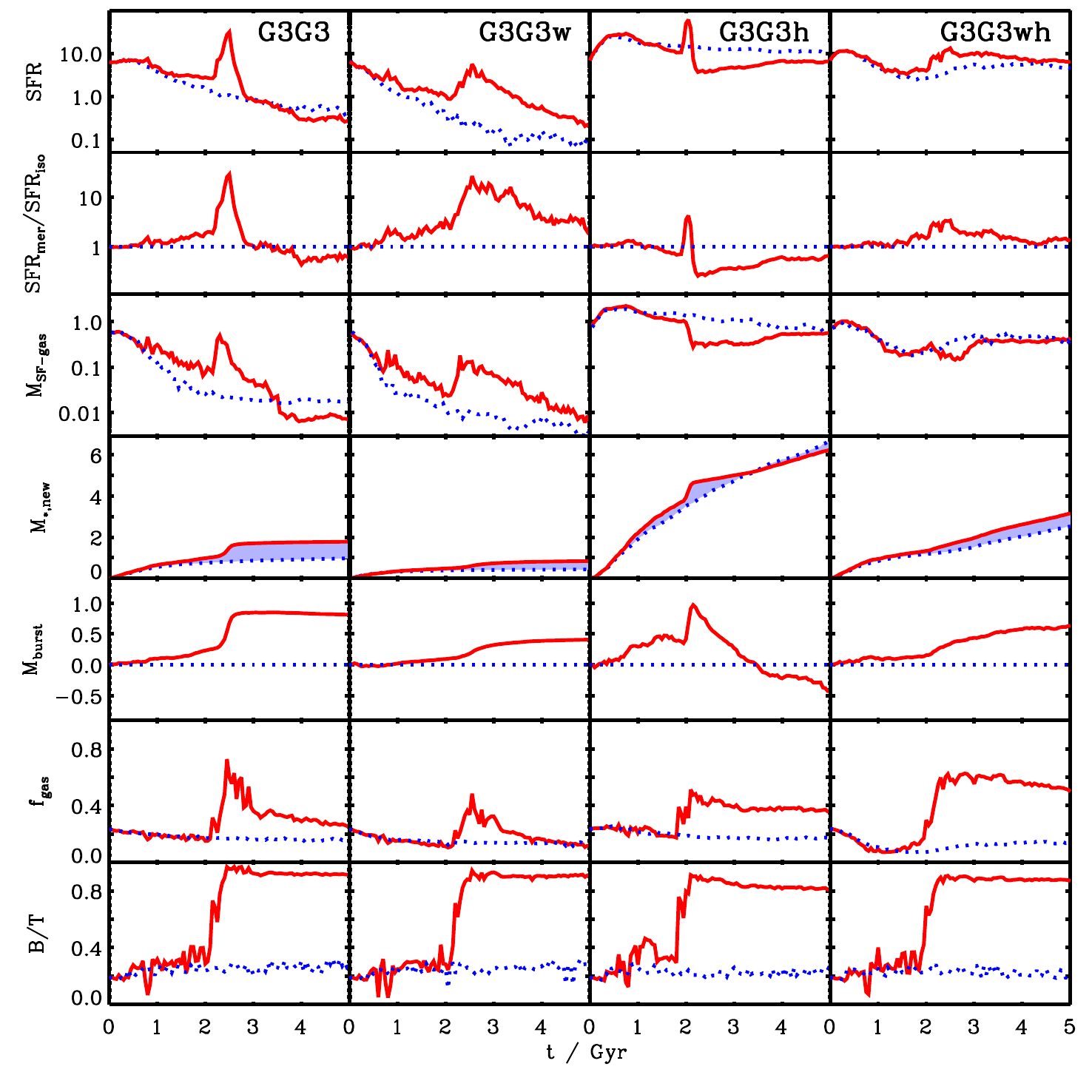,width=0.95\textwidth}
\caption{The rows from top to bottom show the total SFR, SFR
  normalized by the SFR of both isolated galaxies, the mass of dense
  star-forming gas available, the new stellar mass formed during the
  simulation, the burst mass, the gas fraction in the disc and the
  bulge-to-total ratio for the fiducial simulations G3G3, G3G3w, G3G3h
  and G3G3wh, from left- to right-hand side. The results for the
  mergers are shown by the solid red lines and for the combined isolated
  galaxies by the dotted blue lines. All SFRs are given in $\Msunpyr$ and
  all masses in $10^{10}\Msun$.}
\label{fig:sfr}
\end{figure*}

The discs become tidally distorted as they reach first pericentre
($t\sim0.8\Gyr$), leading to long tidal tails that drive out loosely
bound material while the central regions form bar-like structures. Due
to dynamical friction, orbital energy decreases while the spin of the
haloes increase, which results in an almost radial orbit after the
first encounter. During the second encounter ($t\sim2.3\Gyr$), both
discs are destroyed and the remnant is a spheroidal-looking
object. Due to shocks, some of the kinetic energy of the gas is
transformed to thermal energy and then radiated away, resulting in an
offset between the stellar and the gaseous components. This drags the
gas into the dense central regions where the cooling time is very
short, leading to a starburst. In the G3G3 run, heated low-density gas
expands from the central region and forms a (low mass) hot gaseous
halo after the merger \citep[cf.][]{cox2004}. Gas from this halo cools again
and accretes onto the orbital plane where it forms a cold gaseous disc,
subsequently forming a nuclear stellar disc. The G3G3wh run is
similar, however, due to the additional potential of the hot gas
material, the galaxies merge slightly earlier ($\sim150$ Myr). The stellar
remnant is more flattened than in the G3G3 run. In
addition, after the merger there is still a large reservoir of hot gas
left in the halo which has received most of the orbital angular
momentum. This massive hot halo can reform a much more prominent cold
gaseous disc after the merger which is thinner than in the G3G3
run. In both simulations we find several tidal condensations which
form from loosely bound tidal material and consist of gas and new
stars.
 
\subsection{Star formation}
\label{sec:sfr}

In this and the following section we study the effects of stellar
winds and the hot gaseous halo on the SFRs and the efficiency of the
starburst. For this we compare the SFRs and the new stellar mass that
forms during the simulations between the four merger simulations and
for the progenitor galaxies evolved in isolation. We investigate the
effect of including stellar winds but no hot halo, a hot halo but no
winds, and both winds and a hot halo.

The star formation history for each of our four fiducial mergers is
shown in the first row of Figure \ref{fig:sfr}. The simulation without
stellar winds and no gaseous halo (G3G3) agrees very well with the
results of \citet{cox2008}: The SFR of the merger (solid line) is
clearly enhanced compared to the isolated runs (dotted line). The
maximum SFR during the merger ($\sim30 \Msunpyr$) is 30 times larger
than the summed SFR of the isolated discs ($\sim0.5\Msunpyr$), as
shown in the second row of Figure \ref{fig:sfr}. The starburst starts
shortly before the final coalescence and after the merger, the SFR
quickly drops ($\sim 500$ Myr) to the value of two isolated discs. We
find that in both cases the SFR after 3\Gyr~($<1\Msunpyr$) is very low
compared to the SFR at the beginning of the simulation ($\sim
6\Msunpyr$). This is due to the limited amount of remaining cold gas,
which has no source of refuelling. We show the amount of cold star
forming gas (i.e. cold gas which also fulfills the density criterion
$\rho>\rho_{th}$) in the third row of Figure \ref{fig:sfr}. The amount
of dense star-forming gas strongly increases just before the burst,
due to the torques driving gas into the nucleus. This star-forming gas
reservoir is largely consumed in the burst.

The SFR in the simulation with stellar winds included (G3G3w) drops
much more rapidly for the isolated galaxies, as cold gas is expelled
from the disc by the winds. After $\sim3\Gyr$ the SFR has dropped by a
factor of a hundred. During the merger we also find a starburst, with
an enhanced SFR ($\sim6\Msunpyr$) that is $\sim30$ times larger than
the summed SFR of the isolated discs ($\sim0.1\Msunpyr$). With respect
to the simulation with no winds, the absolute value of the SFR during
the starburst is five times lower. The duration of the starburst is
also significantly increased with respect to the windless case. This
result was already demonstrated by \citet{cox2006b}, although using a
different implementation of stellar feedback. The SFR in the merger is
always larger than in the isolated case and equals that of the
isolated discs only at the end of the simulation. The amount of dense
star-forming gas is lower than in the no wind case during the
burst. After the burst, however, there is as much star-forming gas
available as in the G3G3 run, leading to a similar SFR. The reason for
the longer duration of the starburst is that during the starburst, the
wind prevents some of the gas from reaching the dense nucleus of the
galaxy.  This material is then located in a halo around the galaxy and
can subsequently cool and accrete onto the disc, leading to a higher
SFR.

In the run without winds, but with a hot gaseous halo (G3G3h), the SFR
for the isolated galaxy stays relatively constant during the
simulation ($\sim6\Msunpyr$ per galaxy), as the cold gas in the disc
is constantly refuelled through cooling from the halo.  The starburst
in the merger has an enhanced SFR of $\sim60\Msunpyr$, which is a
factor of five times larger than the summed SFR of the isolated
discs. However, after the burst the SFR drops to a value that is lower
with respect to the isolated case ($\sim5\Msunpyr$). The reason for
this is the lower amount of star-forming gas available after the
burst. This means that the cold gas at the centre of the galaxy which
is compressed and quickly used up during the burst, is not immediately
replenished, indicating that there is a process that hinders the
accreting gas from becoming dense and forming stars.

Finally, the SFR for the isolated galaxies in the simulation including
winds and a gaseous halo (G3G3wh) is also relatively constant
($\sim2\Msunpyr$ per galaxy), but lower than in the G3G3h case due to
the stellar winds which remove dense gas from the centre.  Although in
the merger case there is clearly an enhanced SFR, the peak is not as
prominent and the burst is spread over a much larger time interval
than in the other cases. With respect to the isolated case, the
maximum SFR ($\sim13\Msunpyr$) is enhanced by only a factor of
three. After the burst, the SFR decreases again and reaches a value
that is similar to that of the two isolated systems.  The reason for
this is that the amount of dense star-forming gas is the same in both
runs.

In summary we can point out two major effects: (1) In simulations with
stellar winds, the starburst is spread over a much larger time
interval than in the windless case. This is due to the removal of cold
gas from the dense star-forming region in the centre. This material is
ejected into a halo around the galaxy and can then cool and accrete
back to the centre which results in an enhanced SFR. (2) In
simulations that include a hot gaseous halo the enhancement of the SFR
in a merger with respect to isolated galaxies is much smaller than in
systems that do not include a gaseous halo. The reason is the
continuous accretion of cold gas from the halo and the resulting
higher SFR in the isolated discs.  In simulations without this
accretion, most of the gas is already used up by the time of the
merger and the rest is then quickly converted into stars during the
burst resulting in a high SFR. Isolated galaxies, however, have a very
low SFR at the corresponding time, which results in a large difference
between merging and isolated galaxies. If the gaseous halo is taken
into account, the isolated galaxies also have a considerable SFR, such
that the difference between the enhanced SFR during mergers and
isolated systems is no longer as large.

\subsection{Starburst efficiency}
\label{sec:sbe}

In order to quantify the merger-driven star formation it is common to
focus on quantities based on the overall gas consumption rather than
details of the time-dependent star formation history, since the latter
depends on the adopted feedback model. Various authors have made use
of the \lq burst efficiency\rq~$e$ which is defined as the fraction of
cold gas consumed during the merger minus the fraction of cold gas
consumed by the constituent galaxies evolved in isolation
\citep{cox2008,somerville2008a}. This definition is then useful to
predict the additional mass due to the starburst as a function of
initial cold gas mass. An equivalent definition is $e=M_{\rm burst}/
M_{\rm cold}$, where the \lq burst mass\rq~$M_{burst}=M_{\rm
  *,new}({\rm merger}) - M_{\rm *,new}({\rm isolation})$ is the
stellar mass that formed due to the merger and $M_{\rm cold}$ is the
mass of cold gas in the galaxy before the merger. The two main
quantities to determine the burst efficiency are thus burst mass and
the cold gas mass. It is important at which time these quantities are
defined: the cold gas mass has to be defined just before the final
coalescence where the gas will lose angular momentum. The burst mass
has to be defined just after the merger, i.e. when the SFR of the
merging system is at the same level of the isolated galaxies again,
otherwise it depends on the amount of time that has elapsed after the
merger.

\subsubsection{Dependence on winds and gaseous halo}

\begin{figure*}
\psfig{figure=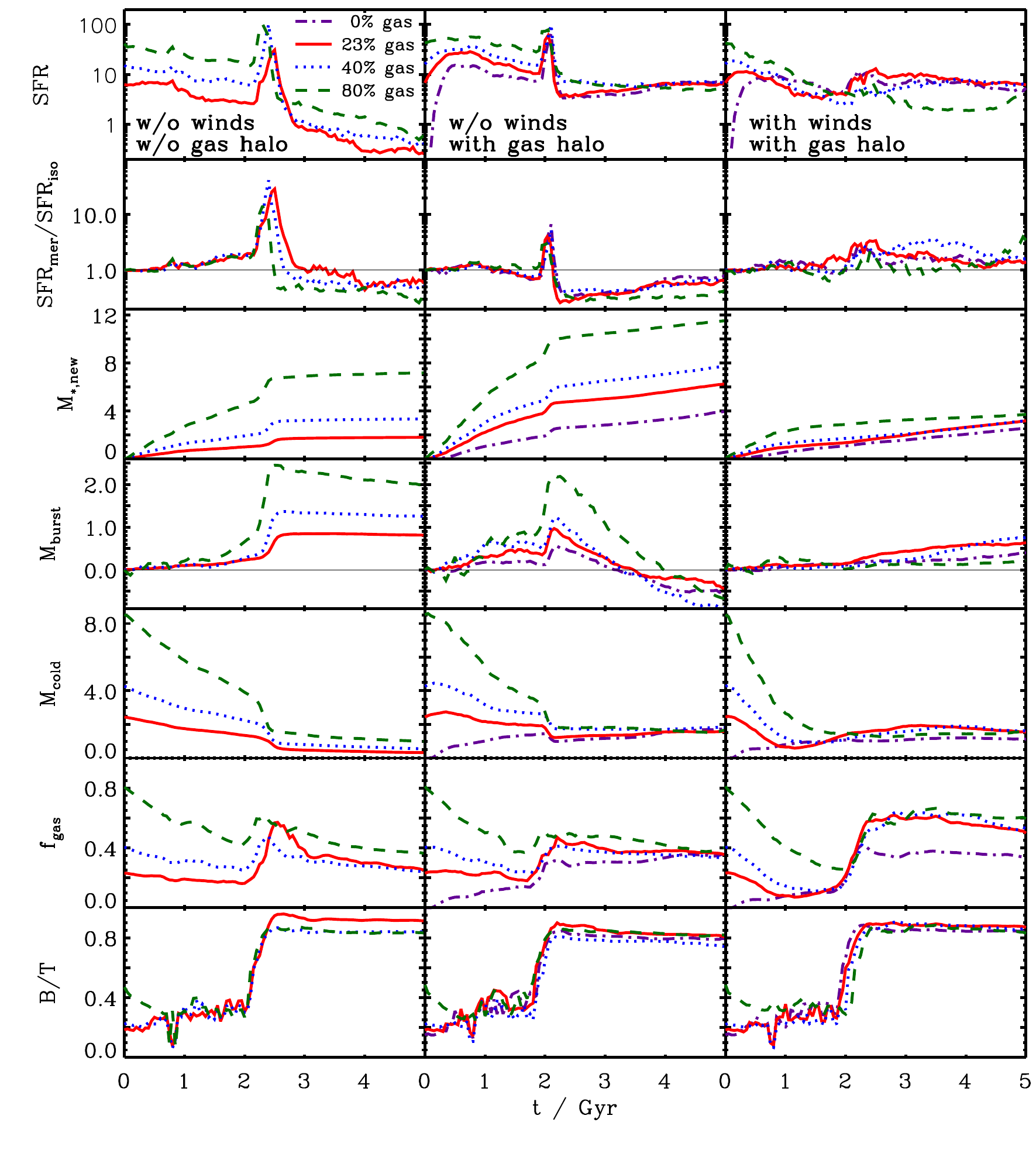,width=0.95\textwidth}
\caption{The rows from top to bottom show the SFR, the SFR normalized
  by the SFR of both isolated galaxies, the new stellar mass formed in
  the simulation, the burst mass, the cold gas mass, the gas fraction
  in the disc and the bulge-to-total ratio for the simulations G3G3,
  G3G3h and G3G3wh from the left- to right-hand side. The various
  lines represent different initial cold gas fractions in the
  disc. All SFRs are given in $\Msunpyr$ and all masses in
  $10^{10}\Msun$.}
\label{fig:sfrgas}
\end{figure*}

In the fourth row of Figure \ref{fig:sfr} we show the new stellar mass
formed during the simulation for the isolated (dotted line) and the
merger case (solid line). The difference between the isolated and the
merger lines is shaded. In the fifth row we plot the burst mass and in
the sixth row, we show the cold gas fraction in the disc $f_{\rm
  gas}=M_{\rm cold}/(M_{\rm cold} +M_{\rm *,disc})$.

The new stellar mass formed in the simulation G3G3 is enhanced during
the starburst, such that the burst mass is $M_{\rm burst}
=0.84\times10^{10}\Msun$. With a cold gas mass just before the burst
of $M_{\rm cold}=1.25\times10^{10}\Msun$ (for both galaxies) this
results in a starburst efficiency of $e=0.68$. Due to the stellar
winds, the starburst in the G3G3w case is less efficient: it has a
burst mass of $M_{\rm burst}=0.4\times10^{10}\Msun$ and a cold gas
mass before the burst of $M_{\rm cold}=0.9\times10^ {10}\Msun$ which
results in an efficiency of $e=0.45$. This shows that any
parametrization of the starburst efficiency depends on the stellar
wind model and its parameters.

In the simulation with a gaseous halo and no winds the mass of the
merging system also increases during the burst with respect to the
isolated systems. It has a burst mass of $M_{\rm
  burst}=0.97\times10^{10}\Msun$ and a cold gas mass before the burst
of $M_{\rm cold}=1.92\times10^{10}\Msun$ resulting in an efficiency of
$e=0.51$. The presence of the hot gaseous halo thus reduces the
efficiency of SF, although the absolute SFR is larger in this case. In
addition, as the SFR of the merger remnant is lower than that of the
isolated systems after the burst, the summed stellar mass of the
isolated galaxies increases faster, such that both simulations have
the same stellar mass at $t\sim3.5 \Gyr$. At the end of the
simulations, the summed stellar mass in the isolated galaxies is
larger than that of the merger remnant. The SFR after the starburst is
lower than that of the isolated systems, since the amount of dense
star-forming gas available is a factor of $\sim4$ lower. The initial
cold gaseous disc is consumed during the starburst, so that the SFR
after the burst depends only on the amount of gas that can be accreted
into the dense nucleus. This means that the accretion is more
effective in the isolated case than after the merger.

In order to explain what prevents the gas from cooling and becoming
dense enough to form stars, we compare the specific angular momentum
and the temperature of the hot gaseous halo of the merger remnant and
the isolated system at $t\sim2.5\Gyr$.  The specific angular momentum
of the gaseous halo in the merger remnant is 25\% higher than that of
the isolated system, leading to larger centrifugal barrier. As we
showed in section \ref{sec:amgh}, a larger angular momentum decreases
the accretion rate and the amount of gas that can reach high enough
densities to form stars. While in the isolated case the specific
angular momentum stays constant throughout the simulation, in the
merger run the hot gas aquires angular momentum from the merger
orbit. For our prograde, almost coplanar orbit this results in a
larger specific angular momentum after the merger is complete, leading
to less dense cold gas available at the centre and thus to a lower
SFR.

There is also a second effect that leads to a lower SFR after the
merger. When the two gaseous haloes collide, shocks occur and heat the
gas, transferring orbital energy to thermal energy. At $t\sim2.5\Gyr$,
just after the starburst, the temperature of the halo gas is
$\sim70\%$ higher than in the isolated run. This leads to a cooling
time which is $\sim35\%$ longer, so that the hot gas in the halo needs
more time to cool and accrete onto the disc. This further reduces the
accretion rate and thus the amount of cold gas that is available to
form stars.

This shows that if the hot gaseous halo is considered in merger
simulations, it is possible (depending on orbit, gaseous halo mass,
etc.) that isolated systems may have a higher SFR than merging
systems, leading to larger stellar masses in systems that have evolved
in isolation. This is a serious challenge for the simple recipe that
is widely used in semi-analytic models, in which a merger leads to
enhanced star formation and therefore to remnants that are more
massive than the sum of the constituent galaxies evolved in isolation.

In the simulation with a gaseous halo and stellar winds the mass of
the merging system is increased during the burst with respect to the
isolated systems. The burst mass is $M_{\rm
  burst}=0.64\times10^{10}\Msun$ and with a cold gas mass before the
burst of $M_{\rm cold}=1.45\times10^{10}\Msun$ the efficiency is
$e=0.44$. This value is slightly smaller than that for the case with
winds but no gaseous halo, so the presence of the halo reduces the
efficiency also in the case with winds. After the burst, the SFR of
the merger remnant is still higher than that of the isolated systems
and reaches a similar value at the end of the simulation, such that
the stellar mass of the merger remnant is larger than the summed mass
of the isolated galaxies. In the isolated runs the winds subsequently
remove small amounts of cold gas from the dense nucleus, so that the
overall SFR is lower than in the case without winds.  This material is
redistributed into the hot gaseous halo and accretes back onto the
disc along with the other cooling gas from the hot halo. In the merger
run, a lot of material is removed from the centre during the burst
(since the wind rate is proportional to the SFR) in a short time
interval, such that suddenly there is a lot of extra material in the
hot gaseous halo. This denser and more massive halo is then able to
cool more rapidly, leading to more cold dense gas and thus to a higher
SFR than in the isolated case. This shows that if stellar winds are
included, more stellar mass may be formed in mergers than in isolated
galaxies.

Another effect worth noting is that the stellar mass formed during the
simulations with a gaseous halo is much larger than in the runs
without it.  The final mass of the new stars in the G3G3h (G3G3wh) run
is a factor of 3.5 (3.7) times larger than the mass in the G3G3
(G3G3w) run, which shows that the gaseous halo is very important to
obtain galaxies with realistic masses. In both of the runs with the
gaseous halo, all of the initial cold gas in the disc has been
consumed by the end of the simulation, while in the other two runs
some of the initial cold gas is still left. This left-over gas does
not reach densities that are high enough to form stars. If the gaseous
halo is present, however, the constant accretion replenishes the gas
disc and ensures that the gas reaches densities high enough to form
stars.

\subsubsection{Dependence on gas fraction}

\begin{table}
 \begin{minipage}{0.47\textwidth}
  \caption{Burst efficiency for different gas fractions.}
 \centering
  \begin{tabular}{lrrrrr}
  \hline
  Run & $f_{\rm g, init}$\footnote{gas fraction in the initial disc} & $f_{\rm g,burst}$\footnote{gas fraction in the disc just before the starburst} & $M_{\rm gas}$\footnote{cold gas mass just
  before the starburst in $10^{10}\Msun$} & $M_{\rm burst}$\footnote{burst mass just after the starburst in $10^{10}\Msun$}&
  $e$\footnote{starburst efficiency}\\
 \hline
 \hline
G3G3 & 0.23 & 0.16 & 1.253 & 0.849 & 0.678\\
G3G3f4 & 0.40 & 0.25 & 2.119 & 1.365 & 0.644\\
G3G3f8 & 0.80 & 0.41 & 3.809 & 2.448 & 0.643\\
\hline
G3G3hf0 & 0.00 & 0.14 & 1.371 & 0.553 & 0.403\\
G3G3h & 0.23 & 0.18 & 1.920 & 0.968 & 0.504\\
G3G3hf4 & 0.40 & 0.24 & 2.640 & 1.218 & 0.461\\
G3G3hf8 & 0.80 & 0.38 & 3.580 & 2.220 & 0.620\\
\hline
G3G3whf0 & 0.00 &0.11 & 1.048 & 0.400 & 0.382\\
G3G3wh & 0.23 & 0.11 & 1.446 & 0.636 & 0.440\\
G3G3whf4 & 0.40 & 0.11 & 1.287 & 0.672 & 0.522\\
G3G3whf8 & 0.80 & 0.26 & 1.405 & 0.224 & 0.159\\
\hline
\label{t:sfeff}
\end{tabular}
\end{minipage}
\end{table}

Up to now we have studied mergers with just one fixed initial gas
fraction in the disc (23\%). In the following section we study how our
results depend on the initial progenitor gas fraction. It has been
shown by \citet{hopkins2009} that the dominant process that removes
angular momentum from the gas during a merger is the lag between the
stellar bar and the gaseous bar. In very gas rich systems, the stellar
mass density is low and the stellar bar is weak or absent. In this
case, as shown by \citet{robertson2006b} and \citet{hopkins2009},
even major mergers may produce little enhancement in star formation
activity and may result in a disc-dominated remnant. Thus, based on
merger simulations with no winds and no hot halo, we expect the star
formation efficiency in the burst and the remnant morphology to scale
with the initial gas fraction in the progenitors.
For this study we use our fiducial models, and keep the total mass of
the disc fixed but modify the gas content in the disc to 0\%, 23\%,
40\% and 80\%. All other parameters remain unchanged from the fiducial
values.

The results are plotted in Figure \ref{fig:sfrgas} for the simulations
G3G3, G3G3h and G3G3wh from left- to right-hand side. From top to
bottom the SFR, the SFR normalized by the SFR of both isolated
galaxies, the new stellar mass formed in the simulation the burst
mass, the cold gas mass, the gas fraction in the disc and the
bulge-to-total ratio are shown as a function of time. In order to
compute the burst efficiency, we use the cold gas mass and gas
fraction just before the starburst, and the burst mass just after
it. The resulting quantities are given in Table \ref{t:sfeff}.

For the case without winds and gaseous halo (G3G3) we find that the
starburst efficiency decreases slightly with increasing gas fraction.
This means that if more gas is available in the disc, it is harder to
convert all of the gas into stars during the burst. If a hot gaseous
halo is included, however, the starburst becomes more efficient with
increasing gas fraction. In order to explain this it is important to
state that the efficiency mainly depends on the amount of cold gas
available for the starburst. This material is of course not the total
mass of cold gas in the simulation, but only the cold gas in the
nucleus, where the burst occurs. Thus the efficiency is a strong
function of the spatial distribution of the cold gas.

In the G3G3 series, the fraction of the cold gas at the centre
($<5\kpc$) is very similar in all runs which means that the amount of
gas available for the starburst scales with the total gas mass. In the
G3G3h series, however, we find that the fraction of the cold gas at
the centre is higher for the runs with a large initial gas
fraction. For these runs, most of the dense initial gas disc is still
present when the burst occurs and can participate in it, while for
runs with a lower gas fraction, a large fraction of the initial gas
disc has already been consumed, so that much of the cold gas is
material that has been accreted from the halo. This accreted material
is distributed in a more extended configuration and cannot contribute
to the starburst, as it is far from the nucleus. The result is that
systems with a high initial gas fraction have a larger amount of
material that can participate in the burst than systems with a low
initial gas fraction, such that they are more efficient. The
efficiency of runs with gaseous haloes, where some of the cold gas was
accreted from the halo and is not centrally concentrated, are always
lower than for the corresponding runs without gaseous haloes, where
all the cold gas resides in the more centrally concentrated initial
disc.

In the simulations with stellar winds there are three runs with the
same gas fraction just before the merger (11\%), but with a different
burst efficiency which increases with increasing initial cold gas
mass. The explanation is similar to the case with no winds: runs with
a larger initial cold gas mass retain a larger amount of the initial
dense disc, while in runs with a lower initial gas mass, there is a
larger fraction of gas that has accreted from the halo and is not as
centrally concentrated. Thus the mass of the gas that can participate
in the burst is larger in runs with a higher initial gas fraction,
resulting in a higher burst efficiency.

\begin{figure*}
\psfig{figure=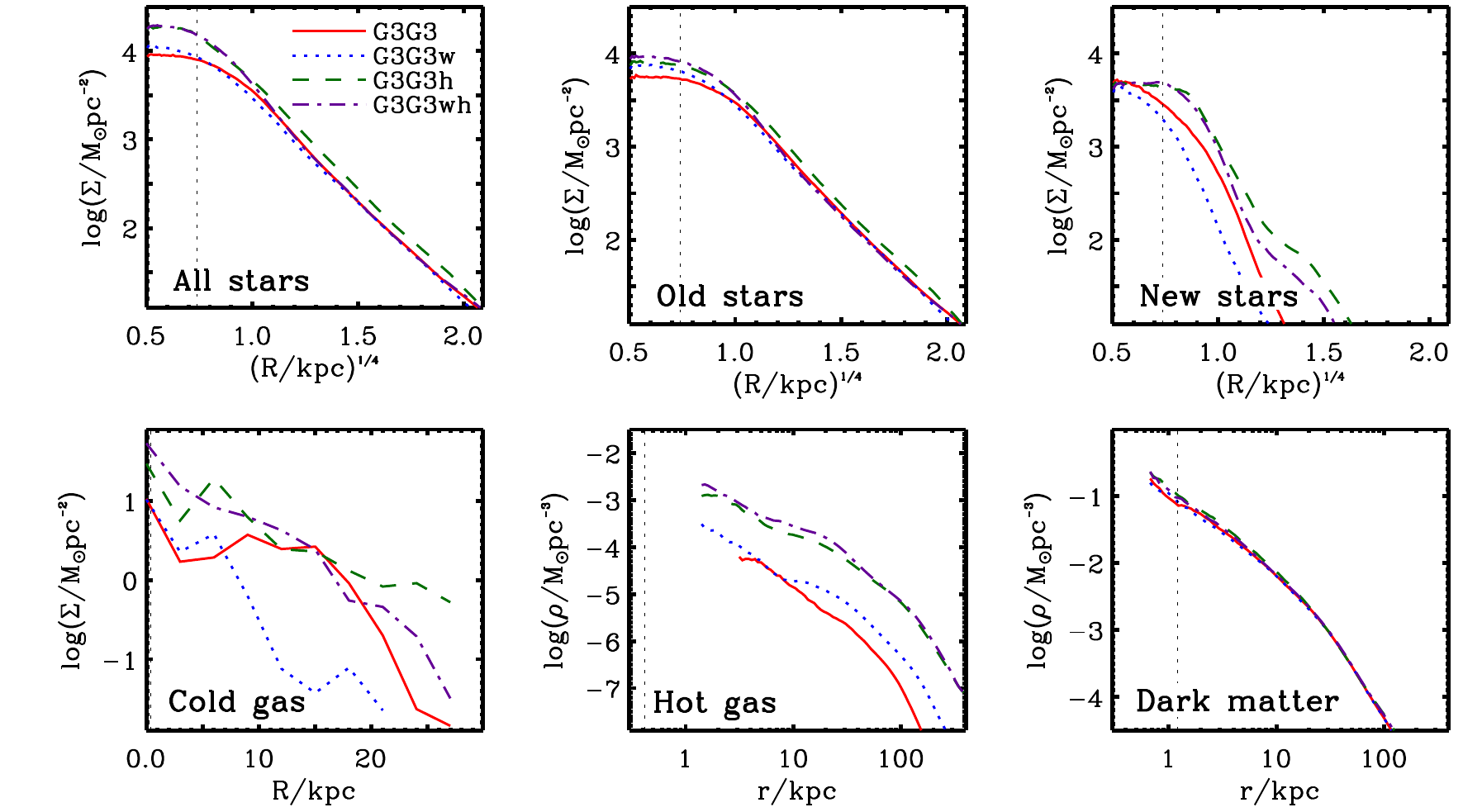,width=0.95\textwidth}
\caption{\textit{Upper panels}: Surface density profiles averaged over
  100 projections for all, old and new stars from left- to right-hand
  side.  \textit{Lower left panel}: Surface density profiles of the
  cold gas as seen face-on. \textit{Lower middle panel}: 3d density
  profiles of the hot gaseous haloes. \textit{Lower right panel}: 3d
  density profiles of the dark matter halo. The various lines show the
  different simulations: G3G3 (no winds, no gaseous halo), G3G3w (with
  winds, no gaseous halo), G3G3h (no winds, with gaseous halo) and
  G3G3wh (with winds, with gaseous halo). The dotted vertical lines
  indicate three times the softening length and hence the resolution
  limit.}
\label{fig:profiles}
\end{figure*}

The run with 80\% initial gas fraction has a lower efficiency than the
other G3G3wh runs, as the burst mass is very low. The main question is
why the run with the high initial gas fraction has a decreased
efficiency with respect to lower initial gas fractions, while for the
simulations without winds the runs with a higher initial gas fraction
have a larger efficiency. To answer this, we measure the mass of the
stellar disc just before the burst: in the G3G3h series the disc mass
is very similar for all runs, as most of the initial gas disc has been
transformed into the stellar disc. In the G3G3wh series, however, the
stellar mass of the disc is very low for high initial gas fractions,
since the stellar winds remove a lot of gas from the disc and thus
lower the SFR. The stellar disc mass of G3G3whf8 is only half the
value of the disc in G3G3wh just before the burst. As a consequence
the amount of stellar mass available to form a bar which can remove
angular momentum from the gas is lower in the G3G3whf8 run, resulting
in less material that can be dragged into the centre during the
burst. Therefore the burst efficiency is reduced.

Overall, we find that the starburst efficiency depends strongly on the
presence of the hot gaseous halo and the stellar winds (and on their
parametrization). When both are neglected, the burst efficiency is a
weak function of gas fraction and decreases with increasing $f_{\rm
  gas}$. Both the gaseous halo and the winds reduce the burst
efficiency. When the gaseous halo is included, the efficiency
decreases with decreasing gas fraction, as the initial dense disc is
replaced by accreted gas from the halo which is spatially much more
extended. When stellar winds are also included, the efficiency for
systems with a high initial gas fraction in decreased, since the winds
prevent the formation of a massive stellar disc before the merger,
which is needed to form a stellar bar that can drag the gas to the
centre of the burst.

\subsection{Morphology of the merger remnant}
\label{sec:mor}

Having investigated how the SFR changes during the merger, we now
focus on the properties of the merger remnant at the end of the
simulation. To this end, we decompose the stellar particles into a
spheroidal and a disc component using the following method. For every
particle, we compute the angular momentum along the spin axis and
divide this by the angular momentum the particle would have on a
circular orbit at the same radius. The spherical component is then
distributed around a value of zero, while the disc component is
centred at unity, such that the two components can be separated easily
\citep{abadi2003}. In the lower panels of Figure \ref{fig:sfr}, we
plot the resulting bulge-to-total ratio $B/T=M_{\rm bulge} / M_*$ for
the four fiducial runs. In all simulations the disc component just
after the starburst is almost completely destroyed.  In the runs
without the gaseous halo, the $B/T$ after the burst is $\sim0.9$ and
stays at this value until the end of the simulation. If a gaseous halo
is included, the $B/T$ after the burst is also $\sim0.9$, but then
decreases to $\sim0.8$ for G3G3h and $\sim0.85$ for G3G3wh. The reason
for this is that due to the accretion of gas from the halo, a much
larger cold gaseous disc can form after the burst which leads to a
much larger stellar disc. Because of the stellar winds in G3G3wh, cold
gas is subsequently removed from the disc, leading to a lower mass
stellar disc at the end of the simulations and thus to a larger $B/T$
compared to G3G3h. Still, all four remnant systems are still very much
bulge dominated.

We also find that the bulge-to-total ratio decreases with increasing
initial gas fraction. However, even for $f_{\rm g,init}=0.8$ the final
value is $B/T=0.85$, while for the lower $f_{\rm g,init}=0.23$ case
the final bulge-to-total ratio is only slightly higher:
$B/T=0.9$. This is mainly due to the low SFR after the burst, so that
the mass in the new stellar disc that forms after the merger is very
low ($M_*=5\times10^9\Msun$ for the 80 percent gas case) as can be
seen in the third row of Figure \ref{fig:sfrgas}. The final
bulge-to-total ratio in the simulations with a hot gaseous halo is
very similar in all runs, independent of the initial cold gas fraction
and whether stellar winds are enabled. The reason for this is that the
SFR after the merger does not depend on the initial cold gas mass and
distribution, but only on the accretion rate after the merger. As the
latter only depends on the mass, temperature and profile of the hot
gaseous component, the SFRs are very similar for all runs, as can be
seen in the first row of Figure \ref{fig:sfrgas}. As a result the mass
in the new stellar disc that forms after the merger is very similar in
all simulations which yields very similar bulge-to-total ratios.

In order to study how the stellar mass is distributed within the
merger remnants, we compute the azimuthally averaged surface density
profiles for old, new and all stars, where old stars are the stellar
particles which were present before the burst (i.e. particles created
for the initial conditions and particles formed through SF) and new
stars are those stellar particles that formed through SF during and
after the burst. The upper panels of Figure \ref{fig:profiles} show
the results for the four fiducial runs. The profiles for all stars at
scales beyond $\sim1\kpc$ are very similar for all remnants following
the observed $r^{1/4}$ profile. The slope is identical for all runs
and only the normalization of G3G3h is higher than that of the other
runs, as the stellar mass before the burst was already much larger.
This can also be concluded from the profiles of the old stars: as the
mass of the old stellar component is very similar for all runs, except
for G3G3h, their final profiles have the same normalization. The
contribution from new stars at these scales is very small for all
runs.
The surface density profiles in the outer parts are in agreement with
profiles obtained by dissipationless simulations \citep{naab2006b},
which shows that the gas physics plays a negligible role far from the
centre.

At small scales ($R\lta1\kpc$) the stellar surface density of the
systems with a gaseous halo is higher compared to the other two
simulations.  Not only is the surface density of the new stars larger,
which can be simply explained by the more massive stellar discs in the
centre, but also the surface density of the old stars is higher. The
reason for this is the central concentration of new stars, which
steepens the potential well and draws the old stars into the
centre. We note that, up to the resolution limit of $\sim3$ times the
softening length \citep[e.g.][]{klypin2001}, the surface density
profiles of the runs with a gaseous halo follow the observed $r^{1/4}$
profile and show neither a cusp nor a core.
Our simulations are in agreement with previous studies \citep{kormendy2009,
hopkins2008,hopkins2009a,hopkins2009b},
who find that including a large fraction of cold gas in merger simulations
leads to a steeper slope in the centre (\lq extra light\rq) and to a lower
S\'ersic index that is distinct from the main body.
However, in contrast to simulations that do not include a hot gaseous halo
we find that large amounts of cold gas are not required in the progenitor
discs, as gas can be accreted from the halo after the merger resulting in
the formation of a dense stellar core.

\begin{figure*}
\psfig{figure=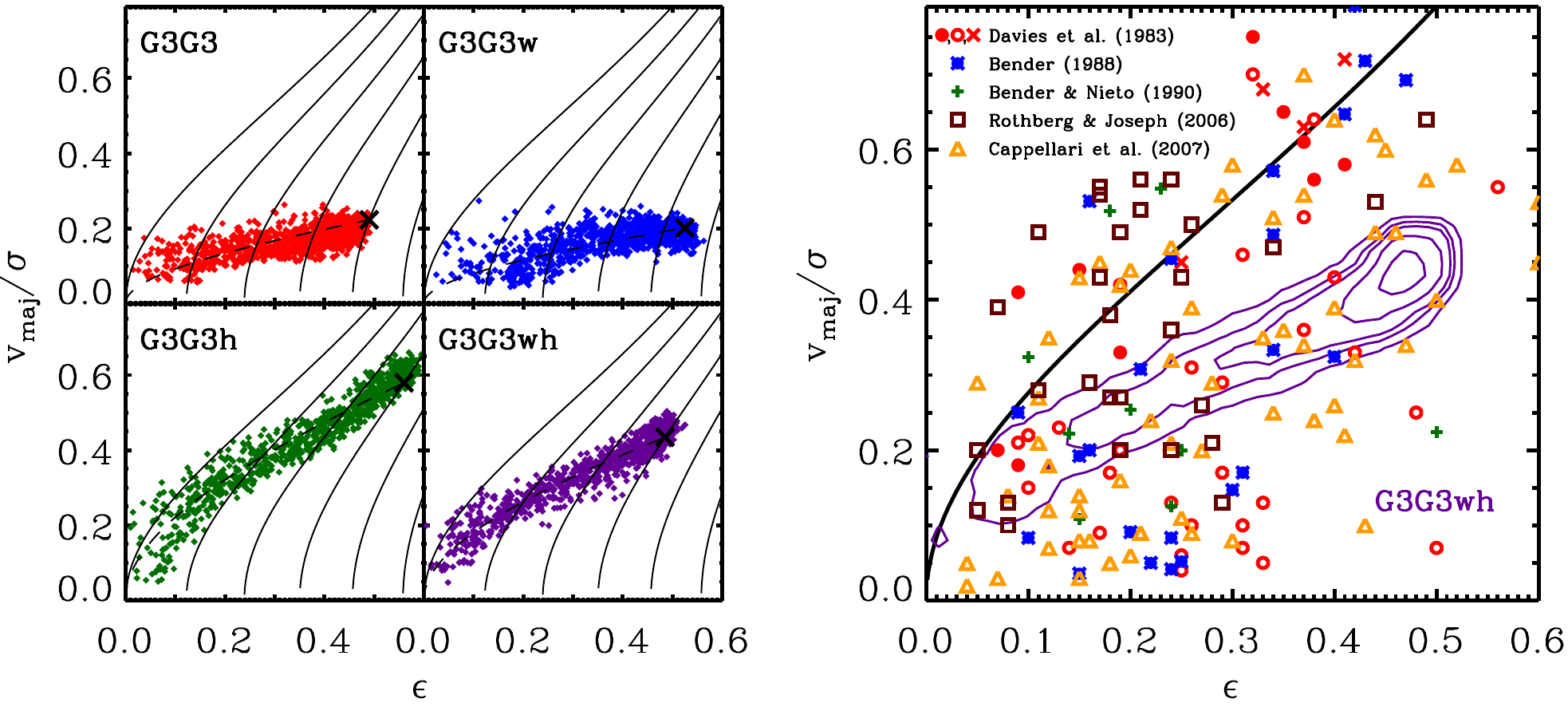,width=0.95\textwidth}
\caption{Anisotropy diagram: $v_{\rm max}/\sigma$ vs. ellipticity,
  where $v_{\rm max}$ is the maximum rotation velocity measured in a
  slit along the major axis and $\sigma$ is the velocity dispersion
  averaged within half of the half-mass radius. The ellipticity is
  measured at the half mass isophote.
  \textit{Left
    panel:} Comparison between the four fiducial simulations, where
  every point corresponds to one projection. The solid lines are the locations expected
  for edge-on oblate galaxies with different anisotropy $\delta=0,0.1,\dots,0.5$. 
  The black symbols indicate the intrinsic ellipticities.
  The dashed lines give the projections for different inclinations.
  \textit{Right panel:}
  Comparison between the simulation with winds and a gaseous halo and
  data from observed elliptical galaxies. The contours indicate the
  10\%, 50\%, 70\% and 90\% probability of finding a merger remnant in
  the enclosed area. The solid line plots edge-on models with $\delta=0$.}
\label{fig:anisotropy}
\end{figure*}

In the lower left panel of Figure \ref{fig:profiles} we show the
face-on density profiles of the cold gas component. In the G3G3 run,
the profile of the gas disc is very flat up to $15\kpc$ but has a very
dense central concentration ($R<2\kpc$). When stellar winds are
enabled much of the cold gas is ejected from the disc, leading to a
very low surface density, except for a dense centre. For the runs with
a gaseous halo, a lot of gas is accreted from the halo and settles
into an exponential disc. The scale length of this disc is lowered if
winds are included. This can be explained by the lower SFR resulting
in a lower stellar mass at the end of the simulation and the fact that
low mass galaxies have a smaller scale length than massive galaxies
\citep{barden2005}.

The lower middle and right panels show the 3d density profiles for the
hot gas and dark matter, respectively. The density of the hot gas for
runs with an initial gaseous halo are almost identical. Only at scales
of $5-50\kpc$, the density in the G3G3wh run is a little higher, as
gas that is heated and ejected from the disc by the winds is added to
the hot gas reservoir and accumulated at these scales.  For the runs
without an initial gaseous halo, we see that a hot halo forms due to
the merger. In the G3G3 run this hot halo forms from gas that is
shock-heated during the merger and has expanded into the halo
\citep{cox_hothalo}. In the G3G3w run, the density of this halo is
enhanced by the material ejected by the wind. The dark matter profile
is nearly indistinguishable for all four runs: an NFW-like profile
modified by the baryonic contraction, with a slightly higher central
density for systems with a gaseous halo.

\subsubsection{Kinematic properties}
\label{sec:kin}

Having determined the mass distribution of the merger remnants, we now
study their kinematic properties, i.e. how much the remnants are
supported by rotation. The elliptical shape of a rotating galaxy can
be understood as a result of flattening induced by
rotation. The elliptical structure of a non-rotating galaxy, however,
must be supported by something other than rotation, most likely by
velocity anisotropy. This trend can be visualized by plotting the
rotation velocity $v_{\rm maj}$ divided by the central velocity
dispersion $\sigma$ against the ellipticity $\epsilon$. In this
anisotropy diagram $v_{\rm maj}/\sigma$ is a measure of the rotational
support and $\epsilon$ indicates the deviation from a circle.

An analytic relation between the intrinsic ellipticity $\epsilon_{\rm intr}$ and
$(v_{\rm maj}/\sigma)_{\rm edge}$ for edge-on projections
has been derived by \citet{binney2005}. For a given anisotropy of the system
$\delta$ one can plot a distinct line in the $\epsilon$--$v_{\rm maj}/\sigma$
plane. For a fixed amount of rotation, the intrinsic ellipticity is lower in isotropic
systems and larger in anisotropic systems. For an observed galaxy at a given
inclination, the observed ellipticity and $(v_{\rm maj}/\sigma)_{\rm obs}$
can be computed from the intrinsic ellipticity and $(v_{\rm maj}/\sigma)_{\rm edge}$
\citep[see][\S4.3]{binney1987}. If viewed face on ($\epsilon=0$), no rotation can
be observed, while edge-on ($\epsilon=\epsilon_{\rm intr}$) the observed rotation
velocity is maximal.

In order to compute the three quantities $\epsilon$, $v_{\rm maj}$ and $\sigma$,
for our simulated remnants, we follow \citet{cox2006}
and refer to this work for details. Here, we give a brief outline of
the procedure. First, the stellar particles are projected onto a plane
as if observed from a random viewing angle. Then we determine the
isodensity contour that encloses half of the stellar mass and fit an
ellipse to it. The ellipticity is computed as $\epsilon=1-b/a$, where
$a$ is semi-major and $b$ the semi-minor axis of the ellipse. A slit
is then placed along the major axis with a length of $3a$ and a width
of $a/4$. This slit is divided into 26 bins, lengthwise, and the
line-of-sight velocity distribution is computed for each bin. Then the
mean velocity and the velocity dispersion are extracted from each
bin. Finally, $v_{\rm maj}$ is defined as the average absolute value
of the maximum and minimum mean velocity along the slit and $\sigma$
as the average dispersion of the bins within $a/2$.

Figure \ref{fig:anisotropy} shows the resulting anisotropy diagram for
our fiducial merger remnants. The left panel compares the four runs
and shows $v_{\rm maj}/\sigma$ vs. $\epsilon$ for 1000 random
projections for each remnant. In each panel we plot the location of edge-on
oblate galaxies with different anisotropy $\delta$ as given by the analytic recipe
(solid lines). For each simulated remnant we compute the intrinsic ellipticity and the
corresponding $(v_{\rm maj}/\sigma)_{\rm edge}$ indicated by the black symbol.
The anisotropy of the remnant can be obtained by finding the line on which
the symbol lies. The dashed lines indicate projections with different inclinations
for the analytic galaxy model and trace the projections of the simulations very well.

The merger simulation with no winds and no gaseous halo leads to a remnant
which is only slowly rotating, with a maximum rotation velocity of $\sim 50\kms$
and a maximum $v_{\rm maj}/\sigma$ of $\sim 0.25$. This means that the G3G3
remnant is mostly supported by velocity dispersion and not by rotation. The
intrinsic ellipticity and the anisotropy parameter are relatively large
($\epsilon_{\rm intr}=0.49$ and $\delta=0.4$). The main
contribution to the rotation comes from the re-formed stellar disc at the centre
of the remnant. This explains why the rotational support increases
with increasing ellipticity, as when viewed edge-on both $\epsilon$
and the observed rotation velocity of the disc are maximal.  When
viewed face-on, one cannot observe any rotation of the disc. For the
simulation with winds, the stellar disc is much smaller, leading to an
even smaller disc contribution to the rotation. This results in lower
values for $v_{\rm maj}/\sigma$ and an anisotropy diagram which is
similar to that of a remnant in a dissipationless simulation. The intrinsic
ellipticity and the anisotropy parameter are even larger ($\epsilon_{\rm intr}=
0.53$ and $\delta=0.45$) than in the simulation without winds.

When a gaseous halo is included, $v_{\rm maj}/\sigma$ increases
strongly with increasing $\epsilon$, which can be explained by the
more massive stellar disc that forms after the merger. When viewed
edge-on, the stellar disc dominates within the slit in which the
velocity is measured and thus leads to a higher mean rotation
velocity. The remnant has an intrinsic ellipticity of $\epsilon_{\rm intr}=0.56$
and corresponds to a galaxy model with an anisotropy parameter of
$\delta=0.3$. This indicates that the elliptical shape of the
remnant at the half-mass isophote ($R\sim5\kpc$) is partly due to
rotation. If stellar winds are included in the simulation, the mass of
the stellar disc is lower, leading to a smaller contribution to the
velocity profiles compared to the run with no winds. This also results
in lower values of $v_{\rm maj}/\sigma$, however, the remnant is still
partly supported by rotation. It has an intrinsic ellipticity of
$\epsilon_{\rm intr}=0.49$ and an anisotropy parameter of $\delta=0.3$
This shows that in the presence of a
cooling hot gas halo, rotationally supported systems can be created
even for moderate initial cold gas fractions, and for orbits which
lead to systems that are supported only by velocity dispersion when the
gaseous halo is not included.

\begin{figure*}
\psfig{figure=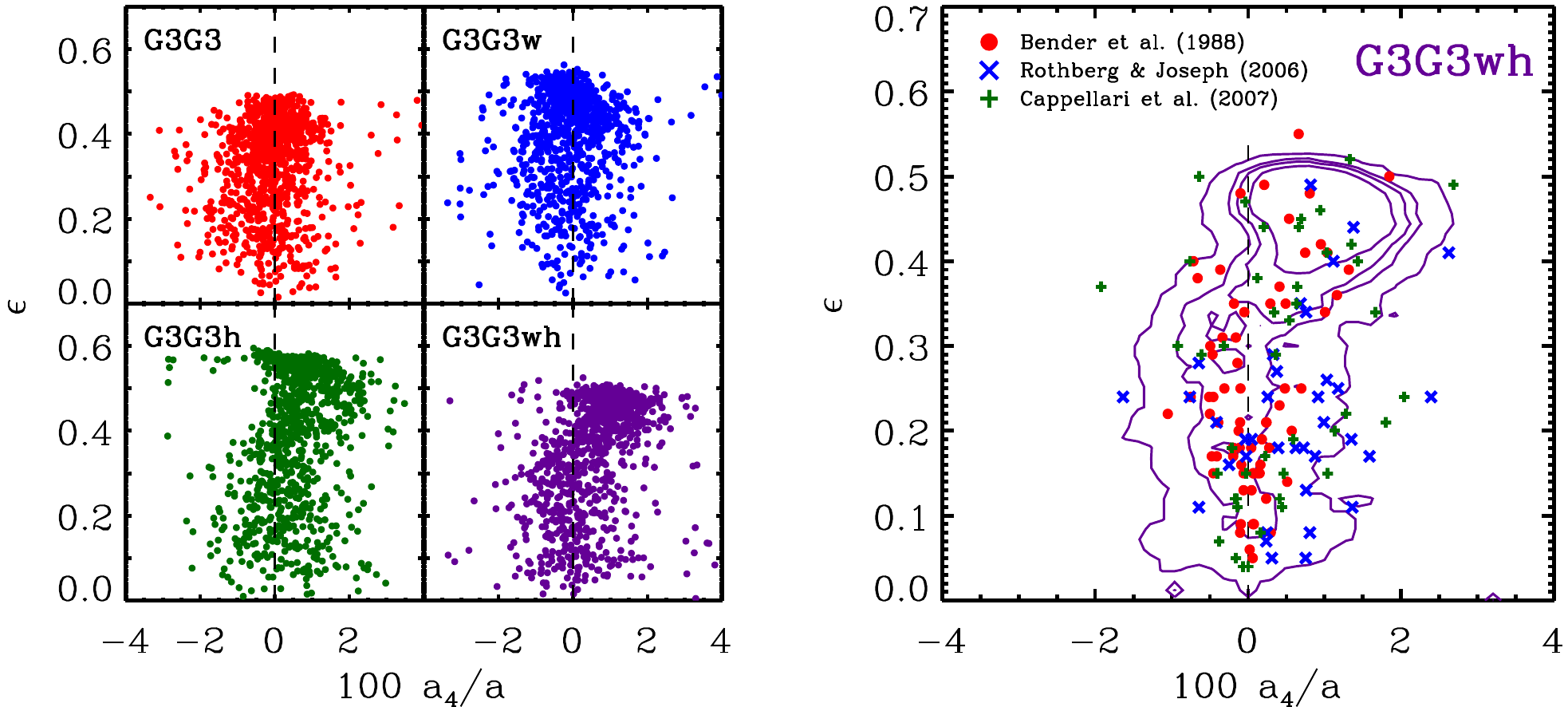,width=0.95\textwidth}
\caption{Ellipticity $\epsilon$ vs. characteristic shape parameter
  $a_4$, both measured at the half mass isophote. \textit{Left
    panels:} Comparison between the four fiducial simulations, where
  every point corresponds to one projection. \textit{Right panel:}
  Comparison between the simulation with winds and a gaseous halo and
  data from observed elliptical galaxies. The contours indicate the
  10\%, 50\%, 70\% and 90\% probability of finding a merger remnant in
  the enclosed area.}
\label{fig:a4}
\end{figure*}

The right panel of Figure \ref{fig:anisotropy} compares the remnant of
the G3G3wh run to data from several observational samples
\citep{davies1983,bender1988a,bender1990,rothberg2006,cappellari2007}. The
contours give the 10\%, 30\%, 50\%, 70\% and 90\% probability of finding a
G3G3wh merger remnant in the enclosed area. It is apparent that the
simulated remnant agrees well with the observations.  There are two
main possibilities that can change the area of $v_{\rm
  maj}/\sigma-\epsilon$ space that the simulated remnant can cover.
First, by decreasing (increasing) the wind efficiency, we can achieve
larger (smaller) values of $v_{\rm maj}/\sigma$ for a given
$\epsilon$, up to the limit of no winds (G3G3h), thus reproducing the
observed remnants with a large rotational support. Second, by lowering
the initial mass of the gaseous halo, we can achieve smaller values of
$v_{\rm maj}/\sigma$ for a given $\epsilon$ and can thus reproduce the
observed remnants with a small rotational support. Note that by
employing the universal baryonic fraction in the initial conditions we
have used an upper limit for the mass of the gaseous halo. Through
feedback, it is possible to decrease the mass of the halo before the
merger. See section \ref{sec:halfhalo} for a discussion.

\subsubsection{Isophotal shape}
\label{sec:shape}

Finally, we study how the shape of the isophotes change when a hot
gaseous halo is included in the merger simulations. To this end, we
measure the deviations of the isodensity contour of a projection to a
fitted ellipse. The residuals are expanded in a Fourier series and the
Fourier coefficient $a_4$ is computed. Positive values of $a_4$
indicate discy isophotes, while negative values indicate boxy
isophotes. The results of this analysis are shown in Figure
\ref{fig:a4}, which plots the ellipticity against the shape parameter
$a_4$ (normalized by the semi-major axis and multiplied by a factor of
100) for our fiducial simulations with 1000 projections each. The left
panel compares the four runs. The simulations with no winds and no
gaseous halo can appear discy or boxy, depending on the viewing angle,
with a tendency towards boxy isophotes. If winds are included, the
isophotes can still be discy, however, they are more boxy on
average. When a gaseous halo is included, the shape of the isophotes
is more discy, especially at high ellipticities. If both winds and a
gaseous halo are present the isophotal projections are mostly
discy. However, at low ellipticities ($\epsilon<0.3$) the probability
for boxy and discy isophotes is equal.

As a result, we find that including a gaseous halo results in more
cold gas at the centre, leading to more discy isophotes. This is in
agreement with results by \citet{springel2000}, \citet{cox2006} and
\citet{naab2006}, who find that mergers with gas increase the fraction
of discy isophotes.  \citet{naab2006} argue that the reason for the
lack of boxy projections in simulations with gas is the different
behaviour of minor-axis tube orbits (which are the dominant family
around the half-mass radius) in axisymmetric and triaxial
potentials. Dissipational mergers lead to more axisymmetric
potentials, and in those, minor-axis tubes look less boxy and more
discy in all projections. Additionally, the fraction of box orbits and
boxlets, which can support a boxy shape, is reduced.
 
The right panel of Figure \ref{fig:a4} compares the remnant of the
G3G3wh run to data from observational samples
\citep{bender1988b,rothberg2006,cappellari2007}. The contours give the
10\%, 50\%, 70\% and 90\% probability of finding a G3G3wh merger
remnant in the enclosed area. The simulated merger remnant agrees very
well with the isophotal shapes of observed ellipiticals. At low
ellipticities, there are both discy and boxy projections, with an
equal probability. At higher ellipticities ($\epsilon>0.4$) the
chances to find boxy isophotes are very small. These trends are
similar to those seen in the observational data.

\subsection{Lower initial hot gas mass}
\label{sec:halfhalo}

In the previous sections we have employed model galaxies with a
maximal hot gas halo, i.e. we chose the hot gas mass such that the
baryonic fraction within \rvir~was the universal value. This was done in
order to demonstrate the maximum effect this hot gaseous halo can have
on the SFR, the starburst efficiency and the remnant morphology, in
comparison to the case in which the hot gas is neglected. Of course,
both cases are the relative extremes, as feedback can decrease the
mass of hot gas in the halo before the merger. Therefore we run a
merger simulation in which we used only half of the \lq full\rq~halo
mass (G3G3whX) as an intermediate case and compare it to the
simulations with the extreme hot gas cases (G3G3w and G3G3wh)
in order to check whether the results of the intermediate case 
fall between the extreme cases. This
additional simulation is evolved for $8\Gyr$ with enabled stellar
winds and using the same orbital parameters as in the previous
runs. 

\begin{figure*}
\psfig{figure= 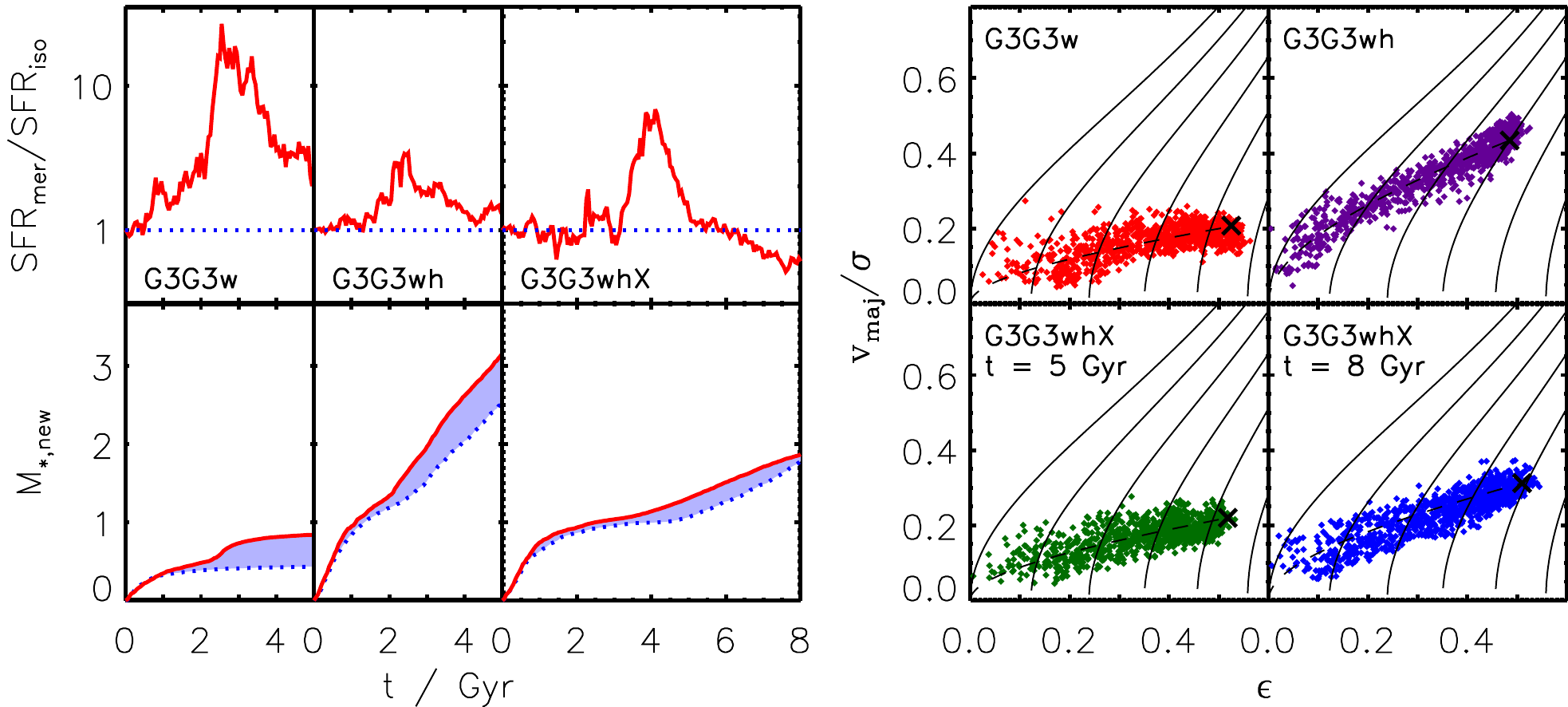,width=0.95\textwidth}
\caption{\textit{Left panels}: The two rows show the SFR normalized by
  the SFR of both isolated galaxies and the new stellar mass formed in
  the simulation for the simulations G3G3w, G3G3wh and G3G3whX, from
  left- to right-hand side. The results for the mergers are shown with
  solid lines and for the combined isolated galaxies by the dotted
  lines. All masses are in $10^{10}\Msun$.  \textit{Right panels}:
  $v_{\rm max}/\sigma$ vs. ellipticity (anisotropy diagrams,
  cf. Figure \ref{fig:anisotropy}). The four panels show a comparison
  between the simulations G3G3w, G3G3wh and G3G3whX at $t=5\Gyr$ and
  G3G3whX at $t=8\Gyr$.  }
\label{fig:halfhgmass}
\end{figure*}

On the left-hand-side of Figure \ref{fig:halfhgmass}, we show the SFRs
during the merger normalized by the SFRs of the respective isolated
galaxies and the new stellar mass formed during the simulations. The
left, middle and right panels show the runs without a hot gaseous halo,
with the \lq full\rq~hot gas mass and with half of the \lq full\rq~hot
gas mass, respectively. The G3G3whX simulations has a maximum SFR that
is enhanced by a factor of $\sim 7$ with respect to the isolated
runs. This value falls just between those of the simulations G3G3w and
G3G3wh with an enhancement of $\sim 30$ and $\sim 3$,
respectively. The new stellar mass that has formed by $t=5\Gyr$ is
$1.3\times 10^{10}\Msun$ for the G3G3whX runs, while in the runs G3G3w
and G3G3wh the stellar mass formed is $0.8\times 10^{10}\Msun$ and
$3.2\times 10^{10}\Msun$.  This shows that the star formation in the
run with half of the \lq full\rq~hot gas mass is an intermediate case
between the two extremes.

With a burst mass of $M_{\rm burst}=0.22\times10^{10}\Msun$ and a cold
gas mass before the burst of $M_{\rm cold}=0.5\times10^{10}\Msun$ the
starburst efficiency for the G3G3whX run is $e=0.44$. This value is
very close to the values obtained for the G3G3w and G3G3wh runs,
$e=0.45$ and $e=0.44$, respectively. This means that in the case with
stellar winds, the starburst efficiency does not depend strongly on
the hot gaseous halo mass.

On the right-hand side of Figure \ref{fig:halfhgmass}, we compare the
kinematic properties of the merger remnants, plotting anisotropy
diagrams for the runs G3G3w, G3G3wh and G3G3whX at $t=5\Gyr$ and for
G3G3whX at $t=8\Gyr$. While the simulation including the \lq
full\rq~hot gas mass forms a remnant that is supported by rotation,
the simulation with half of the \lq full\rq~hot gas mass is
only slowly rotating after $5\Gyr$. However, when evolved to $8\Gyr$, the
rotational support of the system has increased, although not as much as in the
case with the \lq full\rq~hot gas mass. The reason for this is that in
the G3G3whX run, the lower hot gas mass leads to slower cooling and
accretion and thus to a lower SFR. Therefore it takes longer to form a
prominent stellar disc after the merger, and it is this disc that is
responsible for the rotation. As a result we see that the model with
half of the \lq full\rq~hot gas mass can lead to a rotationally
supported remnant as well, although it takes longer and the amount of
rotation is less then for the model that employs the universal
baryonic fraction within \rvir.

Haloes in the Universe have baryonic fractions between the extreme
cases (universal fraction and no hot gas at all). The actual value
will depend on the merger history of a given system, the amount of
feedback during its evolution and its environment. We conclude that
our results for the extreme cases thus give upper and lower limits for
actual galaxies and the results for the simulation with half of the
\lq full\rq~hot gas mass are one possible example within these limits.

%%%%%%%%%%%%%%%%%%%%%%%%%%%%%%%%%%%%%%%%%%%%%%%%%%%
%%%%%%%%%%%%%%%%%%%
%% SECTION 5: CONCLUSIONS AND DISCUSSION
%%%%%%%%%%%%%%%%%%%%%%%%%%%%%%%%%%%%%%%%%%%%%%%%%%%
%%%%%%%%%%%%%%%%%%%
\section{Conclusions and discussion}
\label{sec:conc}

We investigated the role of a cooling gaseous halo in merger
simulations using the parallel TreeSPH-code {\sc GADGET-2}. To do this
we have extended the initial conditions to include a hot gaseous halo
(in addition to a dark matter halo, a stellar bulge and a disc
consisting of stars and cold gas). We adopted the observationally
motivated $\beta$-profile to describe the hot gas and required the
halo to be in hydrostatic equilibrium. Furthermore the gaseous halo is
rotating around the spin axis of the disc. We have fixed the initial
angular momentum of the hot gas by requiring that the specific angular
momentum of the gas halo is a multiple $\alpha$ of the specific
angular momentum of the dark matter halo and treated $\alpha$ as a
free parameter.

Cosmological simulations without cooling, star formation, or feedback
show that the spin parameter of hot gas within dark matter haloes is
the same as that of the dark matter \citep{vdb_halospin}, which would
correspond to $\alpha=1$. However, it has been shown that if this hot
gas collapses to form a disc while conserving its angular momentum,
the disc profiles do not match observations --- there is too much low
angular momentum material \citep{vdb_diskprofiles,bullock2001}. These
results are supported by the results of our simulations: if we allow
hot gas with $\alpha=1$ to cool and form a disc in an isolated halo,
we do not reproduce the observed size vs. stellar mass relation for
discs (our discs are too small). It has been suggested that winds
could preferentially remove low angular momentum material, resulting
in a higher average specific angular momentum for the gas relative to
the dark matter ($\alpha > 1$). Indeed, recent simulations including
stellar driven outflows do show this effect
\citep{governato2010}. However, the results of these simulations are
likely sensitive to the details of the ``sub-grid'' recipes used to
model stellar feedback. We therefore adopted an empirical approach in
order to constrain the initial value of $\alpha$ for our hot haloes.

This was done by modelling a typical MW-like galaxy at $z=1$ and
letting it evolve in isolation up to $z=0$ using different values of
$\alpha$. We employed two observational constraints, stellar mass and
disc scale length, in order to determine the correct $\alpha$. The
results for the control simulation without a hot halo disagreed
drastically with the observations, which supports the importance of
the gaseous halo. Simulations with a low value of $\alpha$ showed both
an increase of stellar mass which was too large and scale lengths that
were too small compared to observations.  Large values of $\alpha$ led
to overly large discs and stellar masses that were too low. Only a value of
$\alpha\sim4$ produced stellar masses and scale lengths that were
within the observational constraints for MW-like galaxies. This value
was then used for all of our simulations with hot haloes.

In order to fix the mass of the gaseous halo we required that the
baryonic fraction within the virial radius should be equal to the
universal value. We made this choice because we wanted to demonstrate
the maximum effects that the hot halo can have in galaxy mergers.  Of
course, due to feedback processes and stellar winds, the baryonic
fraction in galaxies can be lower, leading to a lower mass for the
gaseous halo. In the past, several studies have raised the question of
whether massive gaseous haloes such as those predicted by galaxy
formation models, and adopted in our simulations, violate constraints
from X-ray observations. In one such study, \citet{benson2000}
compared X-ray observations of three massive spirals to predictions of
simple cooling flow models and found that the model overpredicted the
X-ray luminosity by more than an order of magnitude. 
However, they assumed that the hot gas follows either an isothermal
or an NFW profile, which results in very high gas densities in the centre
and such results in very high X-ray luminosities.

Current observational measurements on the $0.3$-$2$ keV X-ray
luminosity of MW-like systems indicate limits of several $10^{39} {\rm
  erg~s}^{-1}$
\citep{wang2003,strickland2004,wang2005,tuellmann2006,li2007,sun2009,owen2009}.
These observational limits have been compared to the X-ray
luminosities found in cosmological hydrodynamic simulations by
\citet{rasmussen2009}. They found no tension between simulations and
observations (for a simulated MW-like galaxy they found $0.3$-$2$ keV
X-ray luminosities of $\sim10^{38}{\rm erg~s}^{-1}$). In a recent
paper \citet{crain2010} showed that disc galaxies in simulations
develop gaseous haloes with an associated X-ray luminosity that is in
good agreement with the observational constraints. They argue that
this lower X-ray luminosity is due to the density profile of the hot
gas not following that of the dark matter and being much less
centrally concentrated as a result of energy injection from SNae.
Similarly, we can compute X-ray luminosities for our assumed hot gas
haloes. To do this we integrate the emissivity over the volume of the
halo in the $0.3$-$2$ keV band (see \citealp{navarro1995} for a
prescription to do this for SPH simulations). For our galaxy models
Z1A4, G3h and MBh we find $L_X = 2, 1{\rm~and~} 10\times10^{38} {\rm
  erg~s}^{-1}$, respectively. We thus conclude that even our maximal
gaseous haloes are not in conflict with X-ray observations.

We have used MW-like galaxy models including a gaseous halo in a
series of binary merger simulations. We have run all mergers both with
and without galactic winds using the `constant wind' model of
\citet{springel2003}, where the mass-loss rate carried by the winds is
proportional to the SFR and the wind speed is constant. However, this
model has some deficiencies as the wind speed for low mass galaxies is
the same as for high mass galaxies resulting in too much heating for
low mass systems. More recent models have been developed that attempt
to overcome these problems and present a more realistic match to
observed quantities, e.g. momentum driven winds
\citep[]{oppenheimer2006}, in which the mass loading factor and wind
velocity are functions of the internal galaxy velocity dispersion.
However, since our study focussed on equal mass galaxies, and our
galaxies do not change their internal velocity dispersions
significantly over the course of the simulation, 
we doubt that introducing such scalings would significantly effect our
results.

We have studied the impact of a gaseous halo and stellar winds on the
SFR using four `fiducial' runs: without gaseous halo and without winds
(G3G3), without gaseous halo but with winds (G3G3w), without winds but
with a gaseous halo (G3G3h) and with both winds and gaseous halo
(G3G3wh). We found that in simulations without a gaseous halo, the
maximum SFR during a starburst is $\sim30$ times larger than that of
the constituent galaxies evolved in isolation. When a gaseous halo is
included, this enhancement is much smaller (factor of $\sim5$). The
reason for this is the higher SFR of the isolated systems when a
cooling halo is included. Isolated systems without a hot halo,
however, consume most of their gas early on in the simulation, such
that the SFR is very low by the time of the starburst in the merger
simulation. We also found that when stellar winds are included, the
starburst is spread over a much larger time interval than in
simulations without winds.  This is because the winds eject cold gas
from the centre of the galaxy into the hot halo.  From there it can
cool and accrete back to the centre resulting in a higher SFR at late
times than in the case without winds.

Furthermore, we studied the effects of the gaseous halo on the
starburst efficiency, which we defined as the stellar mass which
formed due to the burst divided by the mass of cold gas in the
galaxies just before the merger. For the G3G3 run we found an
efficiency of $e=0.68$ while for G3G3w it is only $e=0.45$, showing
that the efficiency depends on the wind model and its parameters. The
presence of a gaseous halo reduces the efficiency in G3G3h to
$e=0.51$. In addition, the SFR after the burst is lower than the SFR
of the constituent galaxies evolved in isolation, resulting in a
remnant stellar mass that is lower than that of the two isolated
galaxies. This occurs for two reasons: first, the specific angular
momentum of the hot gas after the merger is higher than that of the
isolated systems, due to the acquisition of orbital angular
momentum. This leads to a higher centrifugal barrier and thus to a
lower accretion rate and SFR. Second, due to shocks that occur during
the merger, the temperature of the gaseous halo in the merger case is
higher than in the isolated case, leading to a longer cooling time and
a lower SFR. The fact that two non-merging galaxies can have a larger
stellar mass than if they had merged poses a challenge for
semi-analytic models, which generally assume that a merger always
leads to enhanced star formation.

We have also studied how the starburst efficiency depends on the
initial cold gas fraction in the progenitors and found that without a
gaseous halo, it decreases with increasing gas fraction. Higher gas
fractions imply a lower stellar mass, which suppresses the formation
of a stellar bar that can remove angular momentum from the gas
\citet{hopkins2009}. If a hot gaseous halo is present, however, the
starburst efficiency increases with increasing gas fraction, as the
initial dense gas disc is replaced by accreting gas from the halo
which is spatially much more extended. Systems with higher initial gas
fractions retain more of their initial dense gas at the time of the
merger which then leads to a more efficient burst. If winds are also
included, the formation of massive stellar discs in very gas rich
galaxies is prevented such that the starburst efficiency is again
decreased.

Our simulations are closely related to prior studies on the efficiency
of starbursts in binary galaxy mergers, and it is important to compare
the results of our simulations without gaseous haloes or winds to those
obtained by \citet{cox2008} and \citet{hopkins2009}, which also
neglected these effects. Using the same progenitor galaxies (G3) but a
different star formation model, \citet{cox2008} find a starburst
efficiency for G3G3 of $e=0.5$. However, in the definition of $e$,
they have used the gas fraction at the start of the simulation rather
than just before the burst. As the gas fraction decreases until the
time of the burst, the efficiency would increase to $e=0.67$ for our
definition. This is in excellent agreement with our value of
$e=0.68$. \citet{hopkins2009} used a large suite of simulations to
develop an analytic model for the burst efficiency.  For our orbit,
mass ratio and gas fraction this model predicts an efficiency of
$e=0.74$. However, this small difference ($\lta 10\%$) is expected, as
the simulations used by \citet{hopkins2009} to tune their model use a
`softer' EOS ($q=0.25$; see \citealt{springel2005b} for details). As a
result, more gas can lose angular momentum and fall to the centre
during the starburst, increasing the starburst efficiency. 

We have also addressed the question of how the gaseous halo affects
the morphology of merger remnants. Analysing the systems $\sim3\Gyr$
after the final coalescence, we found that when a gaseous halo is
included in the initial conditions, the bulge-to-total ratio is
decreased. This is a consequence of cooling and accretion of gas from
the halo into a cold gas disc at the centre which subsequently forms
stars, leading to a new stellar disc. However, all remnants remain
bulge-dominated ($B/T\sim$ 0.8-0.9). Furthermore, we found that the
presence of a gaseous halo affects the surface brightness profiles of
the remnants. While systems with no gaseous halo deviate from the
observed $r^{1/4}$ profile at small scales, systems with a hot halo
match the observed profile. The reason is the new stellar disc which
leads to a higher surface density at the centre.

A kinematic analysis of the merger remnants showed that if the
progenitor galaxies contain only $\sim20\%$ cold gas and no gaseous
halo, our chosen orbit leads to a remnant which is slowly rotating and
only supported by velocity dispersion. When a gaseous halo is
included, however, the rotational support of the remnant strongly
increases. This can be also explained by the presence of a massive
stellar disc in the centre that forms after the merger. The rotation
of this disc contributes to the total rotational support in a
mass-weighted average and leads to a larger potential at the centre,
which causes `old' stars to fall towards the centre and to increase
their rotational velocity due to conservation of angular momentum. We
have further studied the impact of the gaseous halo on the isophotal
shape of the remnants and found that including a gaseous halo leads to
remnants that are more discy on average. We found that both the
kinematic structure and the isophotal shape of the remnants with a
gaseous halo agree very well with observed ellipticals.

Studying the effects of cold gas fraction in progenitor galaxies on
the kinematic structure of major merger remnants, \citet{cox2006}
conclude that in order to form realistic low-luminosity elliptical
galaxies in merger simulations, the progenitor galaxies must have high
gas fractions of $\sim40\%$. Our results confirm that an ample supply
of cold gas must be present at the centre of the merger remnant, but
suggest that the progenitor galaxies need not have such high gas
fractions. Instead, accretion from a hot gas halo, which is expected
to be present, can provide the necessary supply of cold gas.

As the hot gas in the halo is expected in all progenitor galaxies it has
to be considered in all simulations that aim to reproduce typical observed
galaxies. Our simulations and analysis show that by employing a more
realistic model for the progenitor disc galaxy including this component
it is possible to explain several aspects of observed elliptical galaxies
through equal mass mergers of discs, such as an $r^{1/4}$ stellar surface
brightness profile at small scales and the formation of fast rotators
which dominate the local early-type population \citep[e.g.][]{emsellem2011}.
However, this leads to the opposite problem for massive ellipticals:
if a hot gaseous halo (expected in all progenitor galaxies)
always leads to a rotationally supported system, how can slow rotators
form? Traditionally they have been expected to form only in disc mergers
with mass ratios of 1:1 or 1:2 \citep[see][for a detailed discussion]
{jesseit2009,bois2010}. This is related to the problem of how elliptical galaxies can
have low SFRs and red colours as observed, if a gaseous halo leads to the
continuous accretion of cold gas and thus to ongoing SF. In contrast to
observed elliptical galaxies the remnants in our simulations including the
gaseous halo have relatively high SFRs and young stellar populations.
This general problem has been discussed by \citet{naab2009} who
conclude that elliptical galaxies can only form through a merger of typical disc
galaxies if they have merged more than $3-4\Gyr$ ago and star formation
is quenched afterwards. Moreover, there is evidence that dissipation is important
in the formation of low-mass ellipticals, but less so in massive ones \citep[e.g.][]
{delucia2006,khochfar2006,guo2008,delucia2007,oser2010}. Therefore we need
a process that is more effective at high masses.

In order to solve these problems, there must be a process that prevents the hot
gas in massive haloes from cooling as efficiently as we would otherwise
expect, i.e., a source of heating that is more effective in massive haloes.
There have been various suggestions for what this process might
be, for example, heating by radio jets from AGN
\citep{croton:06,bower2006,somerville2008a}, or gravitational heating
by substructure or clumpy accretion
\citep{khochfar:08,dekel_birnboim:08,johansson2009b}. The former fits in
nicely with the observed correlation between slow rotation, boxy
isophotes, and radio detections.
In addition, mergers may also fuel
\lq bright mode\rq~AGN activity associated with the rapid accretion of
cold gas into the galactic nucleus and onto a central supermassive
black hole. This activity may be able to drive winds that remove cold
gas from the galaxy, resulting in a lower SFR and older stellar populations in the remnant.
However, if star formation is quenched immediately after the merger, then a prominent
stellar disk will not be able to reform, and the remnant will have neither have
\lq extra light\rq~in the centre nor rotation. In this case, fast rotators could still
form from disc mergers with high mass ratios \citep[e.g.][]{barnes1998,naab2003}.
Another possibility is that AGN feedback is delayed,
so that the merger remnant can form a disc before the cold gas is removed.

Moreover, we have not included cosmological accretion from outside the
initial virial radius of the halo, which will be very significant at
high redshift ($z>1$). The radiative cooling model used in our simulation assumes a primordial
mixture of hydrogen and helium and does not depend on the metallicity
of the gas. This could lead to higher cooling rates resulting in larger SFRs
such that even a less massive halo would cool relatively quickly.
We have only simulated binary mergers, although
multiple subsequent mergers are common in the Universe. Furthermore
our simulations only consider one mass ratio (equal mass) and disc progenitors.
Finally our study employs only one orbit.
Thus, a full prediction of the demographics of
early type galaxies (merger remnants) will require a self-consistent
treatment of cosmological accretion, accretion from hot haloes, and
feedback from stars and AGN.

\section*{Acknowledgements} 

% Cui honorem, honorem
%
We thank
Hans-Walter Rix,
Glenn van de Ven,
Arjen van der Wel,
Volker Springel,
Tom Abel,
and Simon White
for enlightening discussions and useful comments on this work.
We also thank Volker Springel for providing the code used
as a basis
to create the initial conditions.
The numerical simulations used in this work were performed
on the PIA and THEO clusters of the Max-Planck-Institut f\"ur Astronomie and
on the PanStarrs2 clusters at the Rechenzentrum in Garching.
BPM thanks the Space Telescope Science Institute for hospitality and
financial support for his visit. BPM also acknowledges a travel grant
from the German Research Foundation (DFG) within the framework of the
excellence initiative through the Heidelberg Graduate School of
Fundamental Physics.

%%%%%%%%%%%%%%%%%%%%%%%%%%%%%%%%%%%%%%%%%%%%%%%%%%%
%%%%%%%%%%%%%%%%%%%
%%  REFERENCES
%%%%%%%%%%%%%%%%%%%%%%%%%%%%%%%%%%%%%%%%%%%%%%%%%%%
%%%%%%%%%%%%%%%%%%% 

\bibliographystyle{mn2e}
\bibliography{moster2011}

\label{lastpage}

\end{document}